\documentclass[manuscript]{aastex}

\shorttitle{Energy release near the PIL}
\shortauthors{Sharykin et al.}

\usepackage{fixltx2e}
\usepackage{float}

\begin{document}

%\title{Onset of Plasma Heating and Electron Acceleration in the Magnetic Field Polarity Inversion Line During Solar Flare of March 15, 2015}
%\title{Onset of Plasma Heating and Electron Acceleration in the Interacting Stressed Magnetic Loops During Solar Flare of March 15, 2015}
\title{Flare Energy Release in the Magnetic Field Polarity Inversion Line During M1.2 Solar Flare of March 15, 2015. Paper I. Onset of Plasma Heating and Electrons Acceleration}

\author{I.N. Sharykin\altaffilmark{1,2}, I.V. Zimovets\altaffilmark{1}, I.I.~Myshyakov\altaffilmark{2}, N.S.~Meshalkina\altaffilmark{2}}

\affil{Space Research Institute of Russian Academy of Sciences (IKI), Moscow, Russia}

\altaffiltext{1}{Space Research Institute (IKI) of the Russian Academy of Sciences}
\altaffiltext{2}{Institute of Solar-Terrestrial Research (ISTP) of the Russian Academy of Sciences, Siberian Branch}

%%%%%%%%%%%%%%%%%%%%%%%%%%%%%%%%%%%%%%%%%%%%%%%%%%%%%%%%%%%%%%%%%%%%%%%%%%%%%%%%%%%%%%%%%%%%%%%%%%%%%%%%%%%%%

\begin{abstract}

We present the study of SOL2015-03-15 M1.2 flare, revealing acceleration of electrons and plasma heating in the sheared twisted magnetic structure in the polarity inversion line (PIL). The scope is to make the analysis of nonthermal electrons dynamics and plasma heating in the highly stressed magnetic loops interacting in the PIL by using X-ray, microwave, ultraviolet, and optical observations. It is found that the most probable scenario for the energy release in the PIL is the tether-cutting magnetic reconnection between the low-lying (3 Mm above the photosphere) magnetic loops within a twisted magnetic flux rope. Energetic electrons with the hardest spectrum were appeared at the onset of plasma heating up to the super-hot temperature of 40 MK. These electrons are localized in a thin magnetic channel with width of around 0.5 Mm with high average magnetic field of about 1200 G. The plasma beta in the super-hot region is less than 0.01. The estimated density of accelerated electrons is about $10^9$ cm$^{-3}$ that is much less than the super-hot plasma density. The energy density flux of non-thermal electrons is estimated up to $3\times 10^{12}$ ergs cm$^{-2}$s$^{-1}$ that is much higher than in the currently available radiative hydrodynamic models. These results revealed that one need to develop new self-consisting flare models reproducing 3D magnetic reconnection in the PIL with strong magnetic field, spatial filamentation of energy release, formation of high energy density populations of nonthermal electrons and appearance of the super-hot plasma.

% Nonthermal gyrosynchrotron emission was modelled using the GX SIMULATOR package and results of the nonlinear force free (NLFFF) extrapolation of the magnetic field.

\end{abstract}

%%%%%%%%%%%%%%%%%%%%%%%%%%%%%%%%%%%%%%%%%%%%%%%%%%%%%%%%%%%%%%%%%%%%%%%%%%%%%%%%%%%%%%%%%%%%%%%%%%%%%%%%%%%%%%%%%%%%%%%%%%%%%%%%%%%%%%%%%%%%%%%%%%%%%%%%%%%%%%%%%%%%%%
\section{INTRODUCTION}

In the standard model of an eruptive two-ribbon solar flare \citep[e.g.][]{Hirayama1974,Magara1996,Tsuneta1997,Shibata2011}, nonthermal electrons are produced due to the magnetic reconnection in the cusp below an erupting plasmoid, causing a coronal mass ejection (CME). Hard X-ray (HXR) emission is generated by nonthermal electrons precipitated into the chromosphere in two sources located in opposite footpoints of magnetic loops under an erupting plasmoid. Soft X-ray (SXR) emission is generated by hot plasma filled flare magnetic loops. In the case of disk observations magnetic field polarity inversion line (PIL) intersects the SXR source and is located between two HXR sources.  Thus, loop-like geometry of the X-ray and microwave emission sources is a usual observational manifestation of the flare energy release process in the frame of the standard model. For example, HXR observations obtained by the Reuven Ramaty High Energy Solar Spectroscopic Imager \citep[RHESSI, ][]{Lin2002} reveal a lot of loop-like HXR emission sources \citep[e.g.][]{Battaglia2005,Veronig2005a,Jiang2006,Guo2012}. Nobeyama Radioheliograph \citep[NORH, ][]{Nakajima1995} observations also often show loop structures in the microwave range \cite[e.g.][]{Kupriyanova2010,Morgachev2014}, although there is an opinion that the real structure can be more complicated \citep[e.g.][]{Warren2002,Zimovets2013,Grechnev2017}.

Classical two-dimensional model of magnetic reconnection assumes the interaction of the opposite-polarity magnetic flux tubes at a null-point. At the reconnection site, plasma is heated and thermal electrons are accelerated forming nonthermal power-law energetic spectrum. However, magnetic reconnection can occur in a magnetic configuration without null points as well \citep[e.g. see for review][]{Priest2002}. For example, twisted magnetic flux ropes (MFR) elongated along the PIL can experience internal magnetic reconnection \citep{Demoulin1996,Gordovskyy2011,Pinto2015}. In such a case, charged particles will be directly accelerated and injected into the reconnected flux tubes \citep{Gordovskyy2011,Gordovskyy2012,Gordovskyy2013,Gordovskyy2014}.

%However there were no study of such scenario. And it is not clear what is the height of reconnecting twisted magnetic flux rope. One need to analyze observations of processes in the PIL during flare energy release.

Another scenario of three-dimensional magnetic reconnection in the vicinity of the PIL is the tether-cutting magnetic reconnection (TCMR). For example, this model was discussed by \cite{Moore2001}, where two crossed magnetic flux ropes interacts in the PIL forming small-scale sheared arcade below the reconnection site. Large-scale erupting magnetic structure above reconnection site is also formed. \cite{Liu2013} demonstrated the possibility of the TCMR in a solar flare using multiwavelength observations and nonlinear force-free extrapolation of the magnetic field. Possibility of a CME triggering by TCMR process was presented in the work of \cite{Aulanier2010}, where numerical MHD modelling was performed.

Another way to trigger energy release in the PIL is to stimulate magnetic reconnection by small scale flux emergence in the PIL. Interaction of the upward moving magnetic flux with the overlaying magnetic field will lead to the current sheet formation with subsequent plasma heating and electron acceleration in the PIL. The numerical MHD simulations of a flare process in the frame of this scenario was demonstrated in the work of \cite{Kusano2012}. There were also some flare observations confirming results of numerical modelling \citep[e.g.][]{Bamba2017,Muhamad2017}.

In the work of \cite{Sharykin2017} flare energy release in the PIL was studied for an M1.2 class solar flare occurred on June~12, 2014. The NLFFF modelling reveals TCMR-like interaction of two magnetic flux ropes with oppositely directed magnetic field in the PIL. The observational results evidence in favor of location of the primary energy release site in the dense chromospheric plasma with strong electric currents concentrated near the polarity inversion line. Magnetic reconnection possibly was triggered by the interaction of magnetic flux ropes forming a current sheet elongated along the PIL. However, there were no strong HXR emission to study the population of accelerated electrons in that flare.

Magnetic reconnection in the PIL can be stimulated in the very initial phase of a flare. For example, there were a lot of observations \citep[e.g.][]{Severnyi1958,Chifor2007,Zimovets2009,Bamba2017a,Wang2017} reporting preflare activity around the PIL in different ranges of electromagnetic spectrum. After the initial phase, the flare can develop following the standard model. Such scenario was discussed recently in the theoretical work of \cite{Priest2017}. According to this work a flare can start with a pre-existing flux rope under an arcade of magnetic loops along the PIL; zipper phase (elongation of flare ribbons along the PIL) is associated with a reconnecting twisted magnetic structure in the PIL. In particular, evidences of the zipper effect, indicating three-dimensional magnetic reconnection along the PIL, were observed in many events \citep[e.g.][]{Grigis2005,Bogachev2005,Liu2009,Qiu2009,Liu2010,Qiu2010,Kuznetsov2016,Qiu2017}.

% These flux ropes are observed as a compact sheared arcade along the PIL in the high-resolution broad-band continuum images from New Solar Telescope.
% However there were no detailed quantitative studies of nonthermal electrons populations in the case of three dimensional magnetic reconnection.

Summing up, it is clear that energy release in the vicinity of the PIL is connected with three-dimensional restructuring of the magnetic field. To our mind there were no detailed quantitative studies of nonthermal electrons in the low-lying magnetic structures elongated along the PIL. Majority of the works devoted to the study of nonthermal electrons considered simple magnetic loop-like geometry in the frame of the standard model to interpret multiwavelength observations of solar flares. Our special interest is to consider a case with highly sheared low-lying magnetic loops interacting with each other in the vicinity of the PIL (i.e. a pronounced TCMR case). In this geometry strong magnetic field component originates along the possible current sheet. The motivation to study energy release in such magnetic configuration is to understand peculiarities of nonthermal electrons population originated during 3D magnetic reconnection and how they are related to plasma heating.

The scope of this work is to make detailed quantitative multiwavelength analysis of nonthermal electrons dynamics and plasma heating in the highly stressed magnetic loops interacting with each other in the close vicinity of the PIL. This aim assumes solution of three main tasks. The first task is to estimate density of electrons accelerated in the PIL and to compare it with density of thermal plasma. It will help to understand how many (in other words, what percentage) electrons are accelerated from the background thermal plasma confined in the magnetic structures in the PIL. The second task is to understand how strong magnetic field in the PIL region where nothermal electrons are appeared and plasma is heated. The third task is to reconstruct magnetic field geometry of the flare magnetic structure in the PIL. To solve our tasks we will use joint observations of X-ray and microwave emissions. We will also analyze optical and extreme ultraviolet (EUV) observations with high spatial resolution to determine fine structure of the flare energy release needed to identify the flare magnetic structure in the PIL and estimate its size. This will help us to compare the magnetic energy released during the flare with the energy contents of accelerated electrons and heated plasma.

The paper is divided into four sections. The first one describes observations (time profiles and images of the flare region selected) made in different parts of electromagnetic spectrum. The second section is devoted to the detailed analysis of X-ray spectra from the RHESSI observations. Flare energetics and estimations of plasma and nonthermal electrons densities are also summarized here. Analysis of magnetic field topology and microwave emission from Nobeyama Solar Radio Observatory is described in the third section. Discussion and conclusions are drawn in the last section.

%%%%%%%%%%%%%%%%%%%%%%%%%%%%%%%%%%%%%%%%%%%%%%%%%%%%%%%%%%%%%%%%%%%%%%%%%%%%%%%%%%%%%%%%%%%%%%%%%%%%%%%%%%%%%%%%%%%%%%%%%%%%%%%%%%%%%%%%%%%%%%%%%%%%%%%%%%%%%%%%%%%%%%
\section{OBSERVATIONS}

%---------------------------------------------------------------------------------------------------------------------------------------------------------------------
    \subsection{EVENT SELECTION}

In this section we describe observations of the solar flare selected for the analysis. There were several criteria of the event selection:

\begin{enumerate}
\item Availability of the RHESSI HXR observations at least up to 50 keV with a sufficient count rate needed for spectroscopy and imaging analysis. X-ray emission sources have to be located close to the PIL in the regions of strong vertical electric currents determined from vector magnetograms. It allows to find the events where nonthermal electrons were transported in the sheared twisted magnetic structures elongated along the PIL.
\item Availability of the NoRH and NoRP observations of the nonthermal microwave emission. Microwave emission sources also have to be located close to the PIL. Joint HXR and microwave observations will allow to make detailed quantitative analysis of nonthermal electrons parameters.
\item Initial flare EUV emission sources from AIA/SDO have to be located close to the PIL. AIA images in the 94 and 335 \AA{} (the less sensitive) channels should not be saturated during the impulsive phase.
\item Flare location is in the central part of the solar disk ($\lesssim 500''$ from the center). Such location is preferable for analysis of HMI vector magnetograms and to minimize the projection effects.
\end{enumerate}

Using these criteria we have found the solar flare of GOES M1.2 class occurred on March 15, 2015, with the start at 22:42 UT and peak at 23:22 UT (according to the GOES data). The flare occurred in the active region NOAA 12297 with the heliographic coordinates S17W38. The flare time profile was composed from three successive subflares. Our study will be devoted to the first one as it satisfies to all criteria listed above and reveal the most intensive HXR and microwave emissions in the vicinity of the PIL, that allows us to investigate accelerated electrons and plasma heating in the PIL. Accordance of the selected flare to the aforementioned criteria will be illustrated in the subsequent subsections.

The available observational data from HMI/SDO allow to make detailed investigation of the magnetic field dynamics in the flare region of the selected event. However, it is out of the scope of this work. The subsequent paper (Paper~II) will describe magnetic data and electric currents in details. This work (Paper~I) is mostly concentrated around X-ray and microwave emissions generated from the PIL region.

%\item Strong photospheric electric currents determined from the HMI vector magnetograms (see Section \ref{SDO}) near the flare emission sources; such currents are necessary for the formation of a twisted magnetic configuration.

%---------------------------------------------------------------------------------------------------------------------------------------------------------------------
    \subsection{FLARE X-RAY EMISSION DETECTED BY RHESSI}

Figure \ref{TP}a shows temporal profiles of the RHESSI X-ray count rates in the $6-12$, $12-25$, $25-50$, and $50-100$~keV energy bands. HXR peaked at around 22:45:46~UT, and the total duration of the impulsive phase (according to the $25-50$~keV data) was about 100~s. Secondary softer HXR peak was around 22:46:27~UT. Time profiles of HXR emission are very similar to the microwave emission from NoRH and NoRP. One can see that 12-25 keV time profile is similar to count rate in the energy band of 25-50 keV. Thus, emission time profiles in these two energy bands did not follow the Neupert effect. Possibly nonthermal component in these energy bands is stronger than thermal one, or plasma heating and nonthermal electrons precipitation to dense solar atmosphere were simultaneous. However, one need to make spectral analysis of X-ray spectra (see the Sec.~\ref{RHESSI_specan}) to make proper investigation of heating and acceleration rates in the flare region.

RHESSI images were reconstructed with the CLEAN algorithm \citep{Hurford2002} using detectors 1,3,5,7, and 8. Detector~1 reveals sufficient count rate modulation to achieve spatial resolution around $2-3''$. Left panel in Fig.\,\ref{jz_RHESSI} shows positions of the HXR 25-50 keV (at the level of 70~\% from the maximum brightness) contours relative to the PIL and vertical electric currents for the three different subsequent time intervals. The PIL and electric currents were calculated from the reprojected onto heliographic grid HMI vector magnetogram. One can see that HXR emission was generated in the region of the strong electric currents. It means that nonthermal electrons were localized in the twisted sheared magnetic structure in the PIL.

In Fig.\,\ref{AIA} and Fig.\,\ref{Hinode} the RHESSI contour maps are compared with the EUV images from AIA/SDO and Ca II images from SOT/Hinode, respectively (subsections~\ref{RHESSI_AIA} and \ref{RHESSI_SOT}). X-ray maps are plotted for two energy bands: 6-12 and 25-50 keV. The first energy range mostly corresponds to thermal emission when the second one is to nonthermal one. All X-ray emission sources were located very close to the PIL in the image plane. SXR emission source in the beginning of the impulsive phase had a worm-like shape, elongated along the PIL. It seems that hot plasma channel was formed. SXR maps in the subsequent two time intervals show the rather compact SXR source located between double HXR sources. Probably, in these cases we observe highly-sheared (about 80 degree, according to Fig4a,c,d) loop-like magnetic structure. Detailed investigation of the spatial structure  of the flare region will be made using observations in other ranges of the electromagnetic spectrum (see the next sections) and analysis of the magnetic field extrapolation results (Sec.\,\ref{nlfff}).

%---------------------------------------------------------------------------------------------------------------------------------------------------------------------
    \subsection{FLARE MICROWAVE EMISSION DETECTED BY NOBEYAMA SOLAR RADIO OBSERVATORY}

\begin{figure}[ht]
\centering
\includegraphics[width=0.75\linewidth]{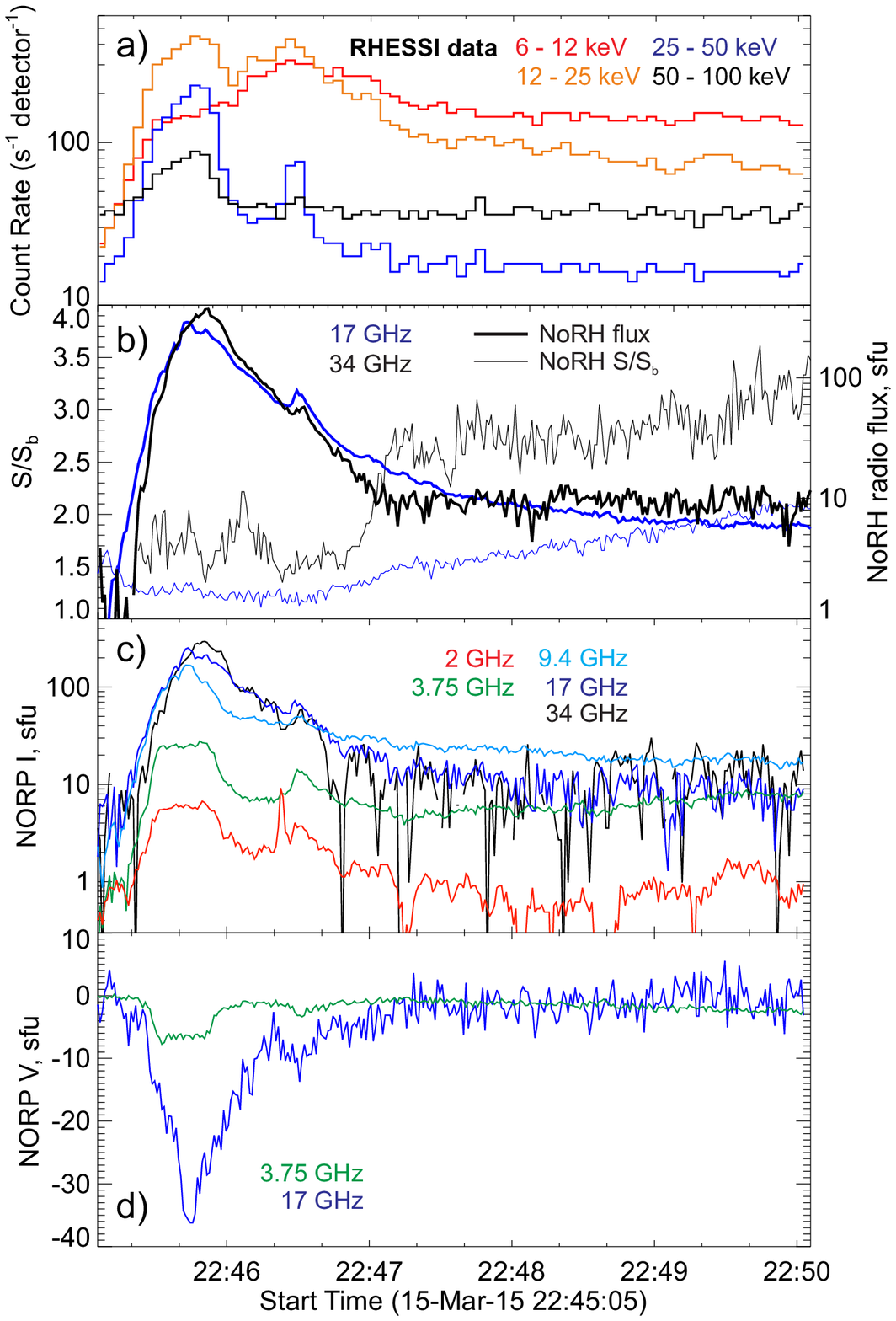}
\caption{Light curves of the M1.2 flare on 15~March, 2015 observed at HXR and radio wavelengths. a) RHESSI count rates ($6-12$, $12-25$, $25-50$, and $50-100$~keV). b) NoRH Stokes $I$ time profiles (17 and 34 GHz) are marked by thick lines. Thin lines correspond to ratio $S/S_b$, where $S$ is the radio source area and $S_b$ is NoRH beam area. c) NoRP Stokes $I$ time profiles at 2, 3.75, 9.4, 17, and 35~GHz. d) NoRP Stokes $V$ time profile at 3.75 and 17 GHz.}
\label{TP}
\end{figure}

\begin{figure}[ht]
\centering
\includegraphics[width=1.0\linewidth]{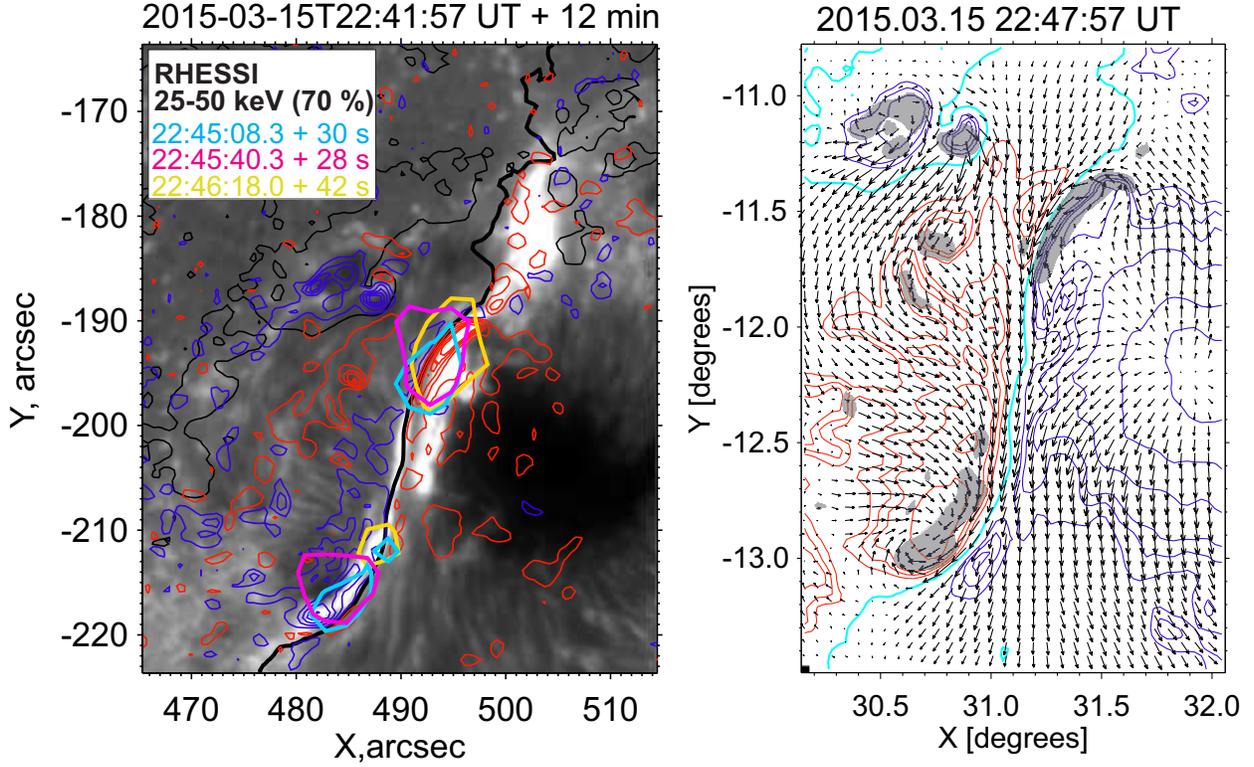}
\caption{Left panel shows comparison of 25-50 keV HXR contours from RHESSI for three different time intervals with Ca II SOT/Hinode cumulative image, contour map of the vertical electric currents and the PIL (black curve) deduced from the HMI vector magnetogram. Red and blue contours (three and five sigma levels) correspond to negative and positive vertical electric currents, respectively. The SOT cumulative image is a result of summing Ca II images taken in time range of 12 min corresponding to the HMI vector magnetogram. Right panel presents the HMI vector magnetogram reprojected onto heliographic grid. Arrows mark horizontal component of the magnetic field. Red and blue contours correspond to vertical magnetic field levels (0.5, 0.75, 1.0, 1.5, 2.0 and 2.5~kG) with negative and positive signs, respectively. Grey regions mark strong electric currents with values above five sigma. Cyan curve is the PIL.}
\label{jz_RHESSI}
\end{figure}

Figure \ref{TP} shows temporal evolution of the selected flare (around the impulsive phase) at different wavelengths. Figures~\ref{TP}b-d show the NoRH and NoRP radio data. NoRP measures both the emission intensity (Stokes~I) and circular polarization (Stokes~V) at five frequencies of 2, 3.75, 4, 9.4, 17 and 34~GHz.  Radio emission peaked at the same time as the HXR (Fig.\,\ref{TP}a), e.g. around 22:45:46~UT. NoRP time profiles also have secondary peak as for the HXR data. Most likely, both radio and HXR emissions were produced by the same population of nonthermal electrons (but in different range of the spectum). The maximum radio flux (at frequency $f_p$) was observed around 35~GHz and was about 350~sfu. Maximal radio fluxes at 9.4 and 17~GHz were a bit earlier. Time delay was about 7~seconds. Radio fluxes at low frequencies was quasi-constant during 25~seconds of the impulsive phase. Peak frequency can not be exactly defined due to low NoRP frequency resolution. However, we can state that $f_p$ was higher than 17~GHz during the HXR and microwave maximum. Such large peak frequency can be explained by the fact that nonthermal electrons with hard spectrum produce gyrosynchrotron emission from the region of strong magnetic field \citep{Dulk1985}. In the Section~\ref{MWanalysis} we will describe analysis of the microwave spectrum in details. The circular polarization was detected only for two frequencies of 3.75 and 17 GHz and the polarization degree was about 25 and 10 \%, respectively. We conclude that the observed microwave radio emission were generated via gyrosynchrotron mechanism.

\begin{figure}[ht]
\centering
\includegraphics[width=1.0\linewidth]{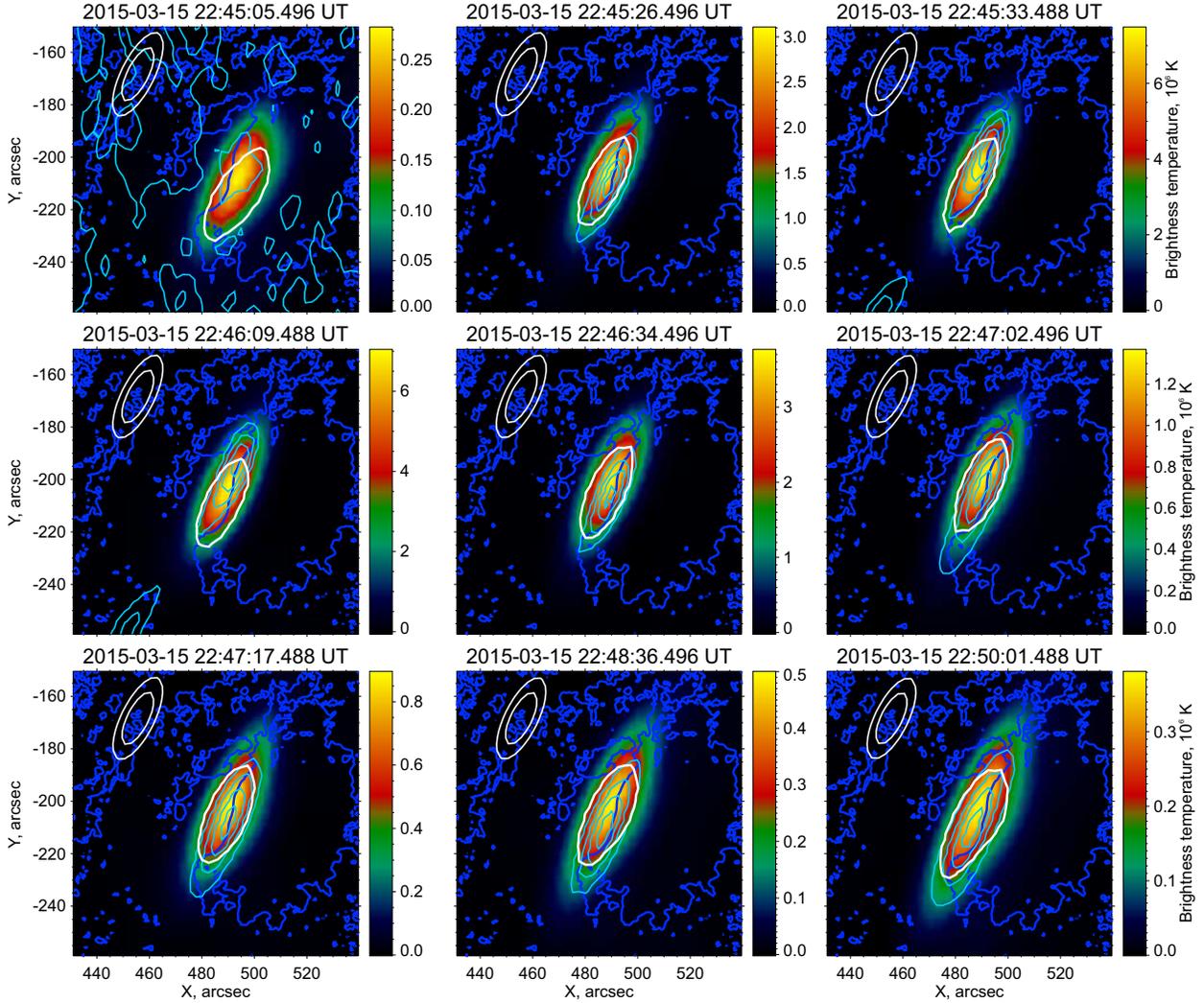}
\caption{NoRH radio images of the M1.2 flare on 15~March, 2015 observed at different time moments. Colored background: 17 GHz Stokes $I$ map; Cian contours: 30, 50, 70, and 90 \% levels of 34 GHz Stokes $I$ map. White contours show negative Stokes $V$ at 17 GHz at FWHM level. Blue line marks the PIL. White large and small ellipses in the upper-left panel corners mark FWHM of the NoRH beams at 17 and 34 GHz, respectively.}
\label{NORH}
\end{figure}

Fig.~\ref{NORH} shows the NoRH microwave images of the flare at nine different subsequent times. NoRH radio flux is presented in Fig.\,\ref{TP}b by thick lines. One can see that NoRH data points are very close to NoRP measurements. NoRH produces microwave maps of the Sun at the frequencies of 17~GHz (Stokes $I$ and $V$) and 34~GHz (Stokes $I$ only) with spatial resolution of up to $10''$ and $5''$, respectively. The images were synthesized using CLEAN algorithm. Due to small elevation of the Sun under horizon resulted NoRH beam was elongated and spatial resolution was reduced up to $15''$ along the large axis of NoRH beam ellipse. One can see that there was only one radio emission source located close to the PIL during all flare time. In Fig.\,\ref{TP}b thin lines mark the time profiles of ratio ($S/S_b$) at 17 and 34~GHz, where $S$ is FWHM area of the apparent source and $S_b$ is FWHM area of the NoRH beam. During maximal radio flux the source size was comparable with the beam size. It means that real emission source was compact. The ratio between source and beam sizes began to increase after 22:47:00~UT and was about 3 for 17~GHz and 2 for 34~GHz at 22:50:00~UT. Thus, emission region experienced expansion during the impulsive phase. This expansion, probably, reflects volume increase of the flare magnetic structures where nonthermal electrons were accelerated and transported.

%---------------------------------------------------------------------------------------------------------------------------------------------------------------------
%    \subsection{HMI/SDO DATA}

%---------------------------------------------------------------------------------------------------------------------------------------------------------------------
    \subsection{COMPARISON OF X-RAY MAPS WITH EUV IMAGES}
    \label{RHESSI_AIA}

\begin{figure}[ht]
\centering
\includegraphics[width=1.0\linewidth]{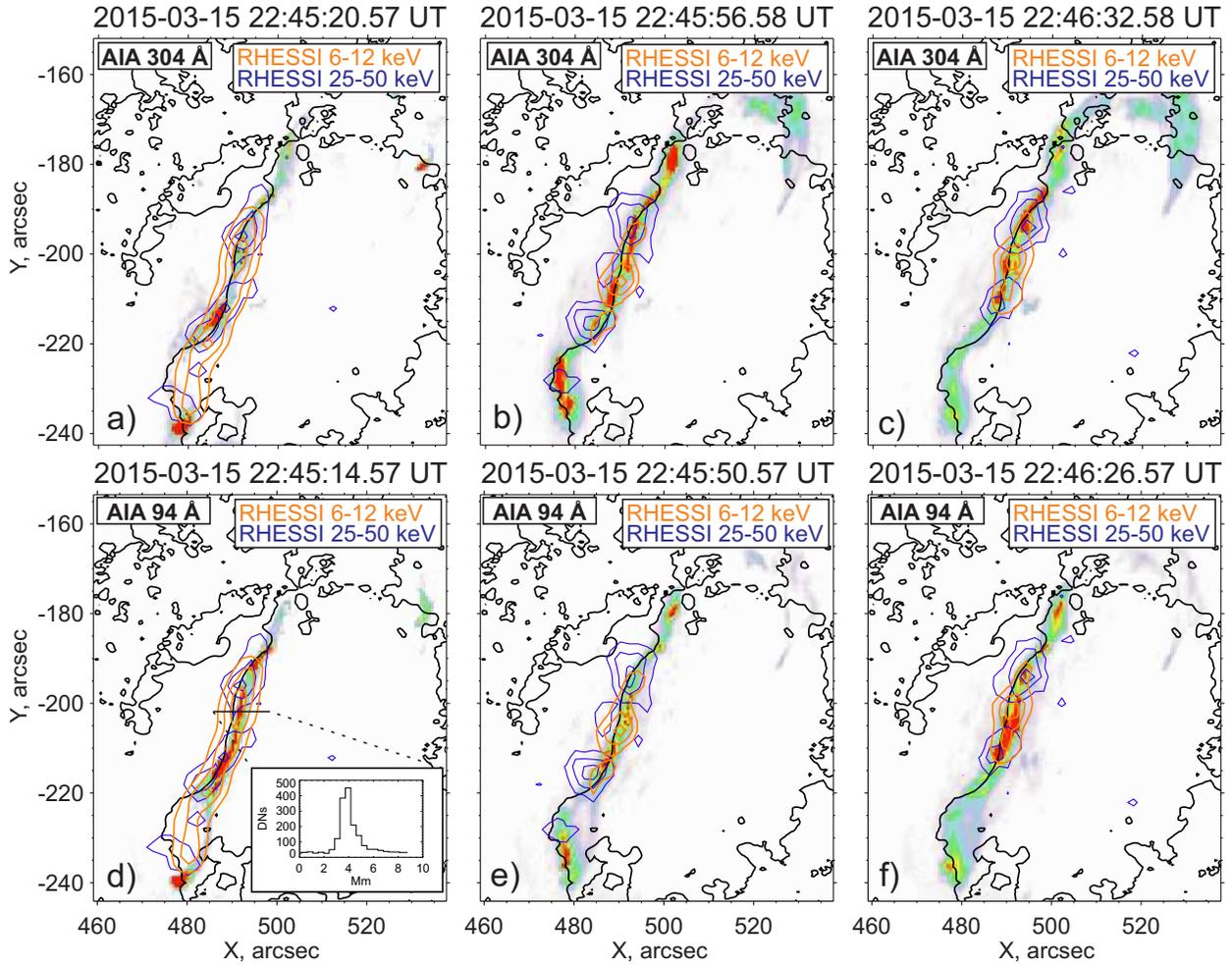}
\caption{EUV maps from AIA/SDO are compared with the RHESSI X-ray contour maps. AIA images in 304~\AA{} and 94~\AA{} channels are shown as colored backgrounds within top and bottom panels, respectively. X-ray contour maps are plotted for two energy bands of 6-12~keV (orange) and 25-50~keV (blue). Image profile along the horizontal slit is presented in the bottom left panel. It allows to estimate the width of the hot flare magnetic structure emitting EUV emission.}
\label{AIA}
\end{figure}

We consider the extreme ultraviolet (EUV) images from two AIA channels of 94 (top panels of Fig.\,\ref{AIA}) and 304~\AA{} (bottom panels of Fig.\,\ref{AIA}), corresponding mainly to the hot ($\sim 10^7$~K) and warm ($\sim 10^5$~K) plasmas. Fig.\,\ref{AIA} shows comparison of the EUV images with the RHESSI X-ray contour maps (6-12 and 25-50~keV) for three time moments. One can notice that all EUV emission sources seen in both channels were very close to the PIL and had the complex spatial structure. The strongest HXR emission was generated in the two HXR sources shown by blue contours in the central part of the image.

During the first considered moment the SXR source had elongated shape with length of $\approx 40''$ along the PIL. The brightest 94~\AA{} emission was generated from the thin filiform source between the two HXR sources and covered by the SXR contour. The width of the observed hot structure is about 2~Mm and was estimated from the image slice shown in the panel. The estimated volume of the hot cylindrical channel is about $9\times 10^{25}$~cm$^{3}$. One also can see the distant compact 94~\AA{} emission source coinciding with the 304~\AA{} source around the point with the coordinates of (-242$''$, 478$''$) where we also observed the weak HXR source. Probably we observe distant footpoint of the flaring magnetic loop inside the MFR.

At the second time interval there were four bright 304~\AA{} emission sources. Two of them are located in the central part of image and correspond to the HXR sources. SXR emission was generated from the compact source located between the two HXR sources. AIA 94~\AA{} image shows the emission distribution similar to 304~\AA{} channel. Probably, we observe the interaction of two sheared magnetic loops in the PIL similar to TCMR interaction \citep{Liu2013}. Four EUV sources could correspond to the footpoints of these loops. In the Section~\ref{nlfff} we will confirm this magnetic geometry by magnetic field extrapolation in the region around the PIL.

One can also note that the brightest EUV emission at the time of the HXR peak (94~\AA{} image at 22:45:51~UT) did not coincide with the location of the SXR (6-12~keV) centroid. The EUV sources on different sides of the PIL are likely hot footpoints of the flare magnetic structures. To explain differences between SXR contours and distribution of the EUV sources from the AIA 94~\AA{} channel one can suppose that RHESSI measures emission from hotter plasma. The maximum of the AIA 94~\AA{} response function is about 7~MK. Temperature in the SXR emitting plasma can be significantly hotter. Thus, we have a very hot coronal magnetic structure observed by RHESSI with ``cooler'' footpoints detected by AIA in the 94~\AA{} channel. However, one need to estimate temperature of the SXR emitting plasma to confirm this. It will be done in Section~\ref{fitting} where the spectral analysis of X-ray spectra from the RHESSI spacecraft is presented.

%---------------------------------------------------------------------------------------------------------------------------------------------------------------------
    \subsection{COMPARISON OF THE X-RAY MAPS WITH SOT/HINODE CA II IMAGES}
    \label{RHESSI_SOT}

\begin{figure}[ht]
\centering
\includegraphics[width=1.0\linewidth]{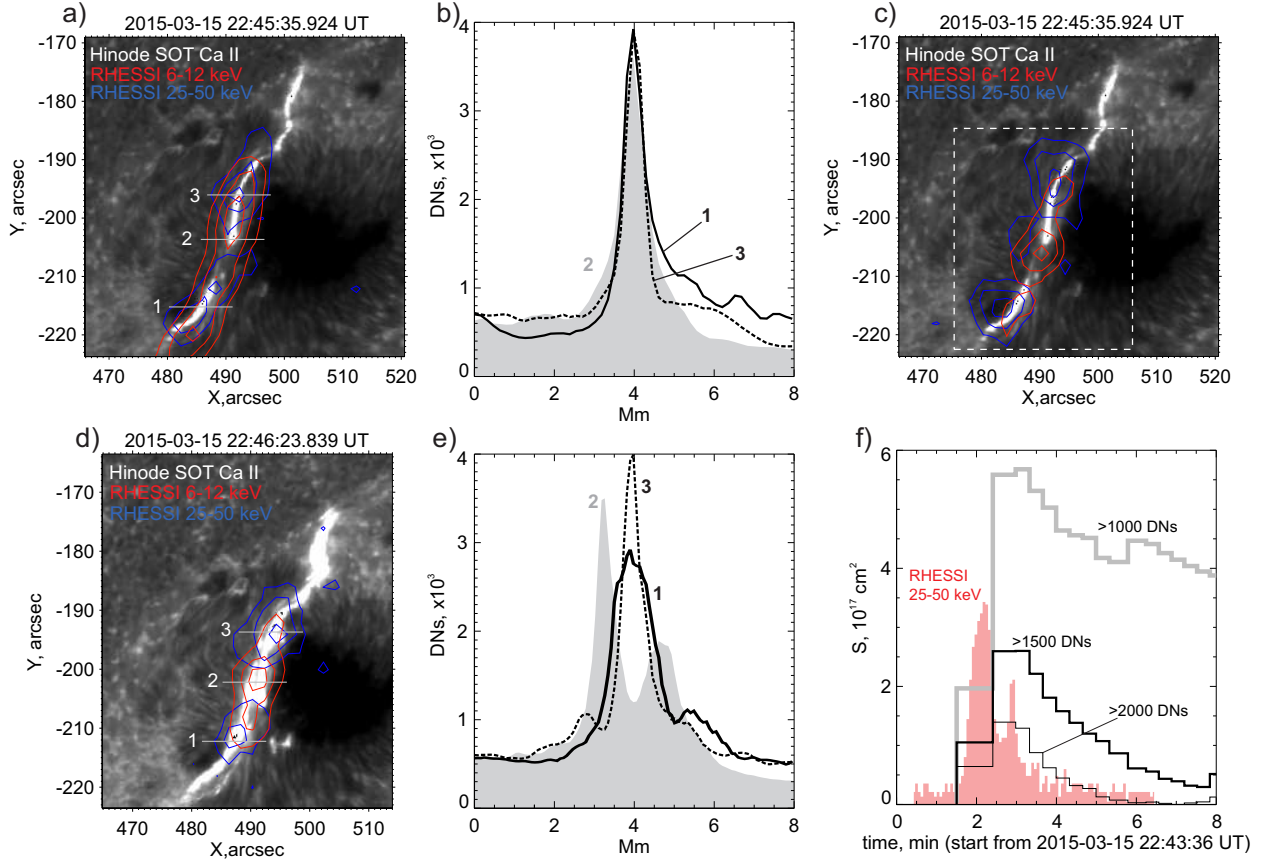}
\caption{Ca~II images (black-white background) from SOT/Hinode are compared with the RHESSI X-ray contour maps in panels~a),~c), and~d). X-ray contour maps are plotted for two energy bands of 6-12~keV (red) and 25-50~keV (blue). Central panels~b) and~e) show slices of Ca~II images made along horizontal observational slits shown in the corresponding a and d panels. These slits were selected to measure the width of the observed flare ribbons. Panel~f) shows time profiles of ribbons area measured for three threshold values of 1000, 1500, and 2000~DNs in the rectangular box shown in panel~d. Red histogram corresponds to RHESSI 25-50~keV count rate.}
\label{Hinode}
\end{figure}

To resolve fine spatial structure of the flare energy release in the lower solar atmosphere we used Ca~II (6684~\AA{}) images from Solar Optical Telescope (SOT: \cite{Tsuneta2008}) onboard space solar observatory Hinode \citep{Kosugi2007}. We have two sets of Ca~II images with different temporal and spatial resolutions covering the flare time period. SOT Ca~II diffraction limited images have the spatial resolution of $0.28''$. The first available set of images have the time cadence of 1~min and pixel size of $0.2''$. The last image of this set at 22:45:36~UT (Fig.\,\ref{Hinode}a,\,c) was made in the rise phase of the HXR and microwave emission. The subsequent array of images has temporal resolution of 20~seconds and pixel size of $0.1''$. The first image of this set was made at 22:46:24~UT (Fig.\,\ref{Hinode}d) and corresponds to the decay phase of the HXR and microwave emissions. In Figure we presented thresholded images to enhance contrast and to demonstrate size and shape of the flare ribbons. Ribbons are thin and intersect sunspot penumbra. One can also see that the flare ribbons were very close to the PIL during the impulsive phase of the studied flare.

We compared the emission sources seen in these two images with the RHESSI X-ray images made in the energy bands of 6-12 and 25-50~keV. The strongest HXR sources coincide with the ribbons seen in the Ca~II images. We see that only part of the ribbon area is covered by the HXR sources. HXR emission is likely to be generated from some particular magnetic loops, whereas the total flare energy release involved larger magnetic structures traced by the optical ribbons.

To estimate the width of the observed Ca~II ribbons we plotted the intensity profiles (Fig.\,\ref{Hinode}b,\,e) along the observational slits marked by short white horizontal lines in Fig.\,\ref{Hinode}a,\,d. There are three positions for observational slits. Slits~1 and~3 intersect only ribbons corresponding to the Southern and Northern HXR sources, respectively. Slit~2 intersects region of the SXR maximal intensity. From these plots it was found that the ribbon FWHM is about 1~Mm. This value is comparable with those one obtained in the work of \cite{Krucker2011}.

Dynamics of area of the flare ribbons is shown in Fig.\,\ref{Hinode}f. To estimate area of the ribbons we considered only the region limited by the rectangular box plotted in Fig.\,\ref{Hinode}c. This box includes the strongest X-ray emission sources. The area was calculated as number of pixels with intensity higher than a threshold limit. There are three threshold values of 1000, 1500, and 2000 DNs. All time profiles revealed that maximal ribbon area was during the second HXR pulse and does not exceed value of $6\times 10^{17}$~cm$^2$. During the HXR and microwave peaks the area was $2\times 10^{17}$~cm$^2$ for threshold of 1000~DNs and $0.6\times 10^{17}$~cm$^2$ for threshold of 2000~DNs.

The obtained values of the ribbon area around HXR sources can be used to estimate lower limit on density of the nonthermal electrons. To do this one can consider nonthermal electrons to be distributed uniformly through all ribbons area. However, it was shown that nonthermal electrons are likely to be injected in local regions of the ribbons. The HXR sources had approximately symmetric shape (let's say circle shape). Assuming the area of nonthermal electrons precipitation to be of circle shape with radius equal to 0.5~Mm (half of the ribbon width) one can also estimate the upper limit for density of the nonthermal electrons. Estimations of nonthermal electrons density will be presented in the next Section~\ref{fitting}.

%%%%%%%%%%%%%%%%%%%%%%%%%%%%%%%%%%%%%%%%%%%%%%%%%%%%%%%%%%%%%%%%%%%%%%%%%%%%%%%%%%%%%%%%%%%%%%%%%%%%%%%%%%%%%%%%%%%%%%%%%%%%%%%%%%%%%%%%%%%%%%%%%%%%%%%%%%%%%%%%%%%%%%
\section{PARAMETERS OF NONTHERMAL ELECTRONS AND HOT PLASMA DEDUCED FROM THE X-RAY SPECTRA}
\label{RHESSI_specan}

\subsection{SPECTRAL ANALYSIS OF THE X-RAY EMISSION}
\label{fitting}

\begin{figure}[ht]
\centering
\includegraphics[width=1.0\linewidth]{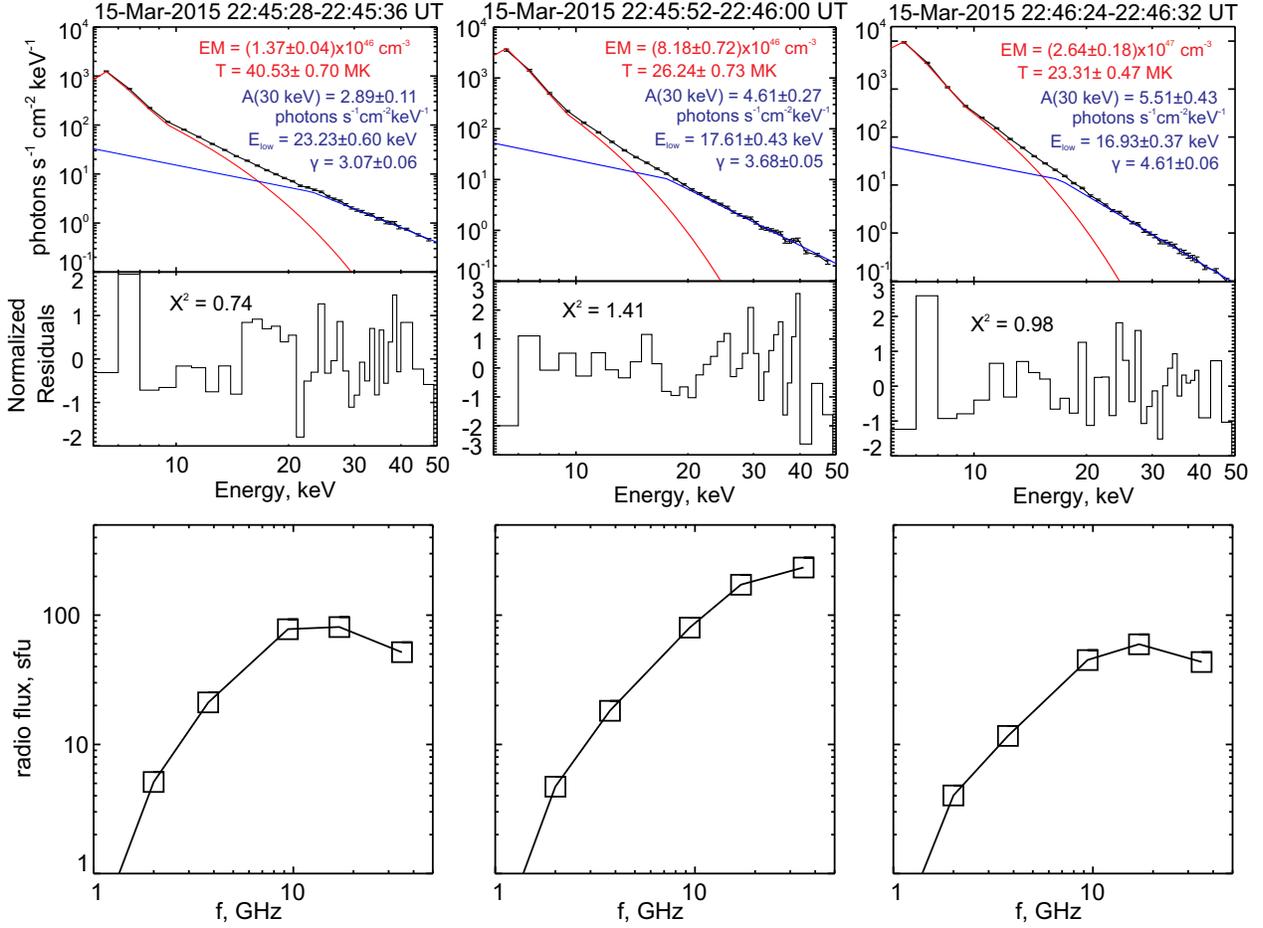}
\caption{RHESSI X-ray (top) and NoRP (bottom) spectra made for the flare impulsive phase. In the top panels, the thermal and nonthermal (double power-law) components of the model fitting function are shown by the red and blue lines, respectively. Parameters of the fittings are written in the plots.}
\label{spec}
\end{figure}

\begin{figure}[ht]
\centering
\includegraphics[width=0.8\linewidth]{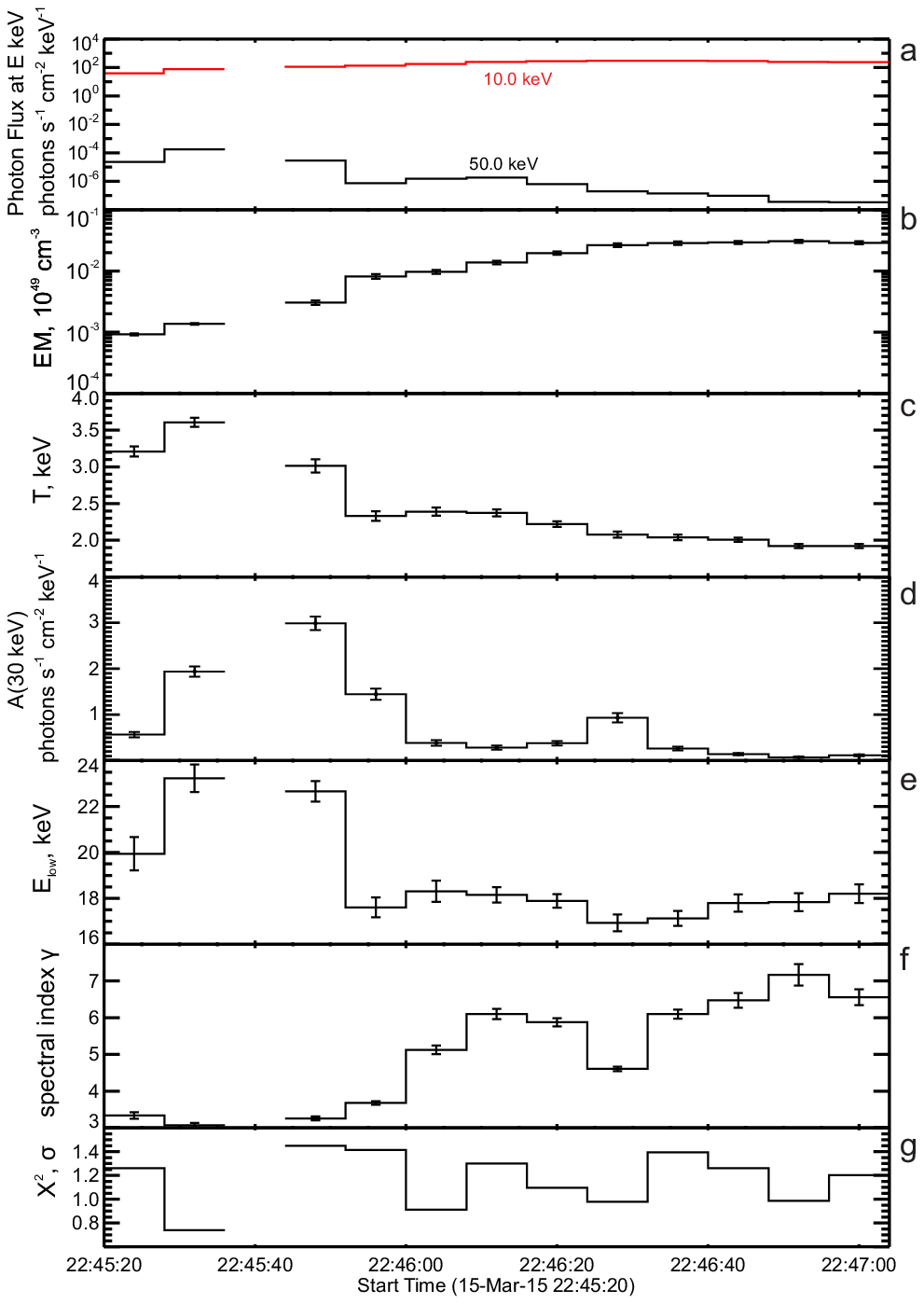}
\caption{Spectral parameters of the X-ray emissions during the flare impulsive phase. a)~Photon flux at 10 (red) and 50~keV (black). b)~Emission measure $EM$ of thermal plasma. c)~Temperature $T$ of thermal plasma. d)~Normalization $A_{30}$ of the HXR spectrum at 30~keV. e)~Break energy $E_{low}$ in the HXR photon spectrum simulating presence of the low-energy cutoff in the spectrum of nonthermal electrons. f)~Power-law spectral indices $\gamma$ of the HXR spectra. e)~Normilized~$\chi^2$ of the fittings.}
\label{fit_res}
\end{figure}

In this section we describe the analysis of the X-ray spectra measured with RHESSI to determine parameters of hot flare plasma and accelerated electrons in the flare region. Figure~\ref{spec} shows the examples of RHESSI X-ray and NoRP radio spectra at three time ranges of the flare considered. The X-ray spectrum (left panel of Figure \ref{spec}) was fitted with a superposition of a single-temperature bremsstrahlung radiation function and a double power-law function to account for the thermal and nonthermal components, respectively. Fitting results are summarized in Fig.\,\ref{fit_res}. Temporal resolution is 8 seconds. A gap in data is due to change of the RHESSI attenuator state.

There are two free fitting parameters for single temperature model of the SXR spectra: temperature $T$ and emission measure $EM$ of the SXR emitting plasma. Contribution of the line emission and free-bound continuum is calculated in the OSPEX using CHIANTI model. Fitting values of $T$ and $EM$ are presented in Fig.\,\ref{fit_res}b,\,c. One can see that there was very strong plasma heating up to 40~MK preceeding to the HXR peak. Such high temperature is referred as super-hot \citep[see ][]{Caspi2010}. Appearance of super-hot plasma can be connected with direct plasma heating in the region close to the magnetic reconnection site. More detailed discussion of the super-hot plasma will be presented in the Section ``Discussions''. Emission measure in the beginning of the impulsive phase was about $10^{46}$~cm$^{-3}$ and increased up to $3\times 10^{47}$~cm$^{-3}$ with plasma temperature value decreased to ``normal'' value of 22~MK.

%To estimate thermal plasma density we use the formula $n_{th} = \sqrt{EM/V}$, where $V$ is volume occupied by thermal plasma. In the beginning of the impulsive phase one can estimate $V$ as $\pi R_{94}*L_{SXR}\approx 8.8\times 10^{25}$~cm$^3$, where $R_{94}\approx 1$~Mm is radii of hot channel elongated along the PIL estimated from AIA 94~\AA{} image (see Sec.\,\ref{RHESSI_AIA}) and $L_{SXR}\approx 28$~Mm is the length of the hot worm-like SXR emission source from the RHESSI image. Considering such geometrical parameters of super hot SXR emitting region $n_{th}\approx 1.1\times 10^{10}$~cm$^{-3}$. At 22:46:20 UT $n_{th}\approx 8.2\times 10^{10}$~cm$^{-3}$ for more compact volume $V\approx 4.5\times 10^{25}$~cm$^3$ (See middle panels in Fig.\,\ref{AIA}) and $L_{SXR}\approx 14$~Mm.

\begin{figure}[ht]
\centering
\includegraphics[width=1.0\linewidth]{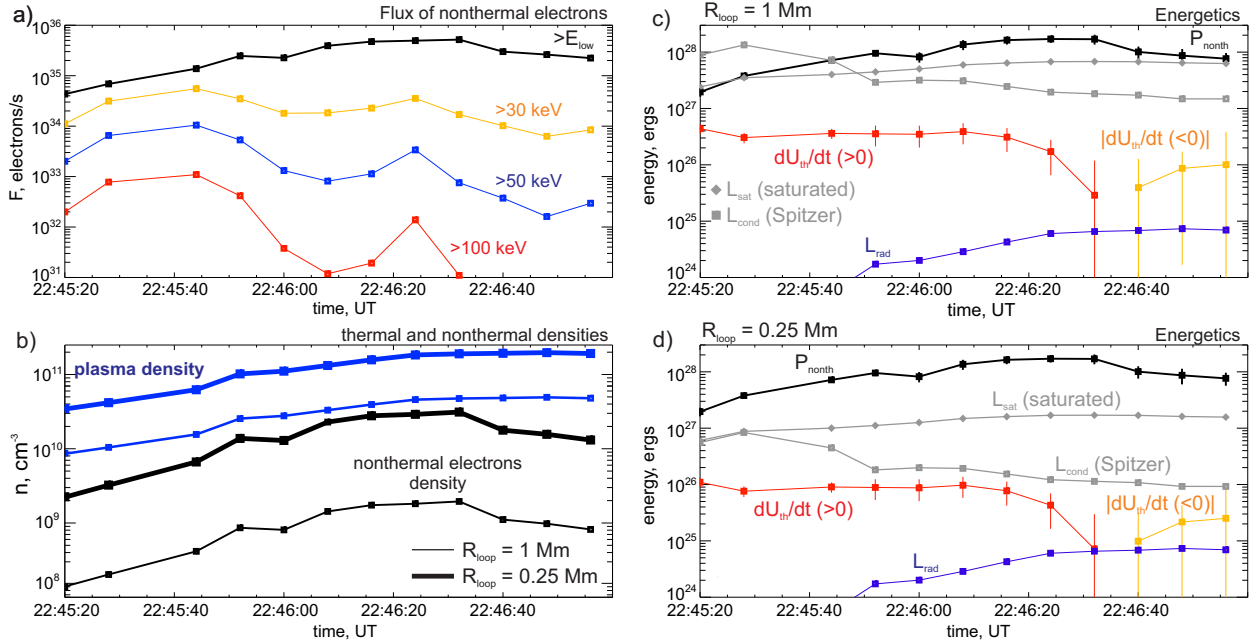}
\caption{Calculation of flare energetics, densities of thermal plasma and accelerated electrons. a)~Fluxes of nonthermal electrons above energies $E_{low}$, 30, 50, and 100~keV. b)~Temporal dynamics of nonthermal electrons density (black) and thermal plasma density (blue). Cases of thin ($R=0.25$~Mm) and thick ($R=1$~Mm) loops are marked by thick and thin lines, respectively. Panels~c) and~d) show comparison of temporal profiles of the rates of different energy channels: kinetic power of nonthermal electrons (black), absolute value of time derivative of plasma internal energy (red and orange), radiation losses (blue), heat conduction losses (grey rectangles), saturated heat flux (grey diamonds). Cases of thin and thick loops are also considered (panels~c and~d, respectively).}
\label{energ}
\end{figure}

To estimate thermal plasma density we use the formula $n_{th} = \sqrt{EM/V}$, where $V$ is volume occupied by thermal plasma. In Fig.\,\ref{AIA} it can be seen that approximate distance between the two strongest HXR emission sources did not significantly change during the flare impulsive phase. One can assume quasi-constant volume of the magnetic loops, where nonthermal electrons were transported and plasma was heated up to super hot temperatures. Moreover, from the NoRH images we know that the microwave sources at 17 and 34~GHz also had quasi-constant area (see. Fig.\,\ref{TP}b,\,\ref{NORH}) during the impulsive phase. That's why, to estimate temporal dynamics of the plasma density we will assume constant volume of the flare magnetic structure. To estimate the volume value we consider loop length as 14~Mm, that corresponds to the distance between the HXR sources. The loop cross-section radius is considered to be 1~Mm; this value was estimated from the AIA 94~\AA{} image (see Sec.\,\ref{RHESSI_AIA} and lower left panel in Fig.\,\ref{AIA}). The time profile of $n_{th}$ is shown in Fig.\,\ref{energ}b by thin line. Thermal plasma density is also calculated for the case of very thin magnetic loop with cross-section radius of 0.25~Mm and shown by thick line in Fig.\,\ref{energ}b.

Nonthermal component of X-ray spectrum ($\gtrsim$20~keV) was approximated by double-power law. The first free fitting parameter is normalization $A_{30}$ (Fig.\,\ref{fit_res}d) of the power-law function at energy of 30~keV. The break energy $E_{\mathrm{low}}$ in the HXR spectrum (Fig.\,\ref{fit_res}e) is also free fitting parameter and simulates presence of the low-energy cutoff in the nonthermal electrons spectrum. The first three time intervals are characterized by $E_{low}=20-24$~keV. Then its value was reduced to $\approx 18$~keV. The low-energy spectral index (at $E<E_{\mathrm{low}}$) of the nonthermal component was fixed at value of 1.5. The third free fitting parameter shown in Fig.\,\ref{fit_res}f is power-law index $\gamma$. Dynamics of the power-law index show soft-hard-soft behavior with minimal value of 3 during the HXR maximum. At the end of the impulsive phase spectrum became the most soft with $\gamma$ up to 7.

The obtained fitting parameters allow us to estimate the total nonthermal X-ray photon flux above $E_{\mathrm{low}}$ as $I_{\mathrm{ph}}(E>E_{\mathrm{low}})=AE_{\mathrm{low}}/(\gamma-1)$ and, then, determine flux of the nonthermal electrons using formula from the work of \cite{Syrovatskii1972}. In Fig.\,\ref{energ}a there are values of the integrated nonthermal electron flux above $E_{low}$, 30, 50 and 100 keV. One can see that the maximal total flux of nonthermal electrons was after the HXR maximum. It can be explained by those fact that at the time of peak flux of the nonthermal electrons spectrum was softer $\gamma\approx 6$ comparing with the time moment of the HXR peak, where $\gamma\approx 3$. The maximal flux of the energetic electrons with energies higher than 30~keV coincided with the HXR and microwave peaks. The flux of accelerated electrons at the HXR peak was $\approx 10^{35}$~electrons/s when its largest value was $5\times 10^{35}$~electrons/s.

The spectral index of accelerated electrons in the HXR source region is related to the emission spectral index as $\delta =\gamma + 1$ using the thick target approximation \citep{Brown1971,Syrovatskii1972}. The nonthermal electron number density $n_{nth}$ (for a power-law spectrum, in the nonrelativistic approximation) is estimated using the formula (with all parameters in CGS units):
\begin{equation}
\label{nnth}
n_{nth}(E>E_{\mathrm{low}}) = \frac{F(E>E_{\mathrm{low}})}{S} \sqrt{\frac{m_{\mathrm{e}}}{2E_{\mathrm{low}}}}\frac{\delta-3/2}{\delta-1},
\end{equation}
where $F(E>E_{\mathrm{low}})$ is integrated nonthermal electron flux above $E_{low}$, $m_{\mathrm{e}}$ is the electron mass and $S$ is the precipitation area of the nonthermal electrons. We estimate the concentration of nonthermal electrons in the case of the thick and thin magnetic loops with radii of 1 and 0.25~Mm, respectively. Values of $n_{nth}$ are shown in Fig.\,\ref{energ}b and compared with thermal plasma density $n_{th}$. The ratio $n_{nth}/n_{th}$ is in the range of 1-2~\% for the thick magnetic loop and $n_{nth}/n_{th}=3-9$~\% for the thin loop. Thus, our estimations show that less than 10~\% of electrons are accelerated from the thermal hot and super-hot plasma population contained in the flare region.

The fit results obtained are used to estimate flare energetics in the next subsection. The parameters of the spectrum of nonthermal electrons are also imported to the GX SIMULATOR \citep{Nita2015} to model the Stokes $I$ microwave spectrum at the NoRP frequencies of 2, 3.75, 4, 9.4, 17, and 34 GHz (see Sec.~4).

\subsection{FLARE ENERGETICS}

In this subsection we will discuss energetics (i.e. the main different energy channels) in the flare region using fitting results from the previous subsection. Internal plasma energy can be calculated following the expression $U_{th} = 3k_BT\sqrt{EM\cdot V}$, where $k_B$ is Boltzman constant. In Fig.\ref{energ}c,\,d we present time derivative of the thermal energy $dU_{th}/dt$ for two cases of magnetic loops of different cross-section radii: 0.25 and 1~Mm (panels~c and d), respectively. Cooling (orange line and dots) of the flare region began approximately after 22:46:35~UT.

Kinetic power of the nonthermal electrons is calculated as:
\begin{equation}
P_{nonth}(E>E_{low}) = F(E>E_{low})E_{low}\frac{\delta-1}{\delta-2}
\end{equation}
The peak value of $P_{nonth}$ is about $2\times 10^{28}$~ergs and was achieved about 22:46:35~UT. This time does not correspond to the HXR maximum as maximal flux of nonthermal electrons was achieved at this time moment. During all flare kinetic power of the nonthermal electrons dominated over the time derivative of the internal energy. In the case of thick magnetic loop with radii of 1~Mm the value of $P_{nonth}/(dU_{th}/dt)\approx 4.5$ in the beginning of the impulsive phase and this value was increased up to 35 in the peak of $dU_{th}/dt$.

In addition to calculating the time derivative of the plasma internal energy and the kinetic power of the accelerated electrons, it is also necessary to take into account the radiative heat losses from the entire super-hot region. For an X-ray-emitting plasma, the heat losses are estimated as $L_{rad}=EM\times10^{-17.73}T^{-2/3}$ for flare temperatures \citep{Rosner1978}. One can see that cooling appeared at the time moment when $L_{rad} \approx dU_{th}/dt$ taking into account errors of $U_{th}$. Close equality of these two energies was achieved in the case of thin magnetic loop with radii of 0.25~Mm.

To estimate heat transfer from the super-hot region to cooler footpoints one can use the assumption of classical (Spitzer) thermal conduction: $L_{cond}\approx 4\times 10^{-6}T^{7/2}/L$, where $L$ characterizes the linear length scale of the temperature gradient that is taken to be equal to the flare loop length. The maximal possible value of the heat flux is estimated as saturated heat flux which is determined by the expression $L_{sat} = v_enk_BT_eS$, where $v_e$ and $T_e$ are thermal electrons velocity and temperature, respectively. This formula means that saturated heat flux assumes heat to be transported along the magnetic loop by thermal electrons spreading with thermal velocity in the same direction. Heat conduction cannot exceed saturated flux. The classical and saturated heat fluxes are presented in Fig.\,\ref{energ}c,\,d by grey lines. Generally, kinetic power of nonthermal electrons also dominates over heat conduction losses. However, in the case of thick loop $L_{cond}>P_{nonth}$ during the first 30~seconds. But taking into account that fact that heat conduction flux cannot be higher than $L_{sat}$ we see that $P_{nonth}\gtrsim L_{sat}$ during entire impulsive phase and both for the case of thin and thick loops considered.

It's worth noting that the maximal thermal energy is usually comparable with nonthermal energy of accelerated electrons \citep[e.g.][]{Emslie2012,Aschwanden2016}. In the case of the studied flare the total kinetic energy of nonthermal electrons in the analyzed time range is about $10^{30}$~ergs. The maximal plasma internal energy for the case of thick magnetic loop did not exceed $3\times 10^{28}$~ergs. We can see two order of magnitude difference between total energies of thermal and nonthermal electrons. However, it is possible that heat conduction from super-hot region could transfer the large part of the released energy to a denser layer of solar atmosphere. Then transferred energy could be radiated in the ultraviolet and optical range of electromagnetic spectrum. In the case of the thick loop the largest possible heat conduction losses considering saturation is about $5\times 10^{28}$~ergs that is 5 times less than total energy of nonthermal electrons and one order of magnitude larger than maximal internal energy. That's why, nonthermal electrons carried the largest fraction of energy in the flare region.

As magnetic free energy (we estimate its value by subtracting potential field energy from NLFFF energy) is believed to be the main source of flare energy one should estimate it as well. The value of the free magnetic energy change during the flare was calculated for the NLFFF (see Sec.~4) extrapolations using two subsequent HMI vector magnetograms (made before and after the flare)  as boundary conditions. The resulted value is about $4.6\times 10^{31}$~ergs that is much larger than one obtained at other flare energy channels discussed above. Thus, the free magnetic energy is enough to explain energy release of the solar flare considered.

It is also worth to estimate density of energy flux $p_{nonth}=P_{nonth}/S$. For the case of thick and thin loops we have values of $1.7\times 10^{11}$ and $2.5\times 10^{12}$~ergs~s$^{-1}$~cm$^{-2}$, respectively. These values are consistent with similar estimates made for X-class helioseismic solar flare of October 23d, 2012, described in the work of \cite{Sharykin2017a}. Thus, nonthermal electron energy flux is much higher than in the currently available flare radiative hydrodynamic models. In the most popular radiative hydrodynamics code RADYN \citep{Allred2006} the maximal considered fluxes was only $10^{11}$~ergs~s$^{-1}$~cm$^{-2}$. One need to make simulations with much higher energy density of nonthermal electrons.

%%%%%%%%%%%%%%%%%%%%%%%%%%%%%%%%%%%%%%%%%%%%%%%%%%%%%%%%%%%%%%%%%%%%%%%%%%%%%%%%%%%%%%%%%%%%%%%%%%%%%%%%%%%%%%%%%%%%%%%%%%%%%%%%%%%%%%%%%%%%%%%%%%%%%%%%%%%%%%%%%%%%%%
\section{ANALYSIS OF RADIO EMISSION}
\label{MWanalysis}

%---------------------------------------------------------------------------------------------------------------------------------------------------------------------
    \subsection{MAGNETIC FIELD EXTRAPOLATION FOR GYROSYNCHROTRON RADIO EMISSION MODELLING}
    \label{nlfff}

To study the magnetic field structure in the flare region, we use observations of Helioseismic Magnetic Imager (HMI, \cite{Scherrer2012}), which provides vector magnetograms with 720 second cadence. We have selected the magnetogram closest to the flare impulsive phase (i.e. at 22:46:00~UT) and recalculated all of the magnetic field $\mathbf{B}$ components from the local helioprojective Cartesian system to the Heliocentric Spherical coordinate system.

To make quantitative analysis of flare microwave emission one need to deterimine distribution of the magnetic field in the flare region. To reconstruct the 3D structure of the coronal magnetic field, we use nonlinear force-free (NLFFF) magnetic field extrapolation \cite{Wheatland2000} with the SDO HMI vector magnetogram used as a boundary condition. The extrapolation is made using the optimization algorithm \citep[implemented by][]{Rudenko2009}.

The extrapolation results are shown in Figure~\ref{GXloops}, where the field lines of the selected magnetic loop are plotted. The side view of these structures are shown in panels~b. It is shown that the height of the bunch of these magnetic field lines did not exceed 3~Mm. Two cases of the magnetic structures are considered. More detailed description of these two variants will be discussed below in the subsection~\ref{GXres}. Generally, the magnetic field lines were chosen to reproduce the observed locations of the HXR footpoints and microwave source. The twisted magnetic structure is elongated along the PIL. The distribution of the magnetic field is not significantly changed along the central line of the magnetic structure. The maximal value is about 1400~G when the minimal one is about 850~G. The resulted magnetic field lines have footpoints located very close to the PIL. This is also in accordance with the observed optical and EUV emission sources in the PIL. Finally, NLFFF modelling reveals the closed low-lying twisted magnetic structure in the PIL, where flare energy release was occured. Thus, plasma heating and acceleration of electrons were stimulated in the found magnetic structure.

\begin{figure}[ht]
\centering
\includegraphics[width=1.0\linewidth]{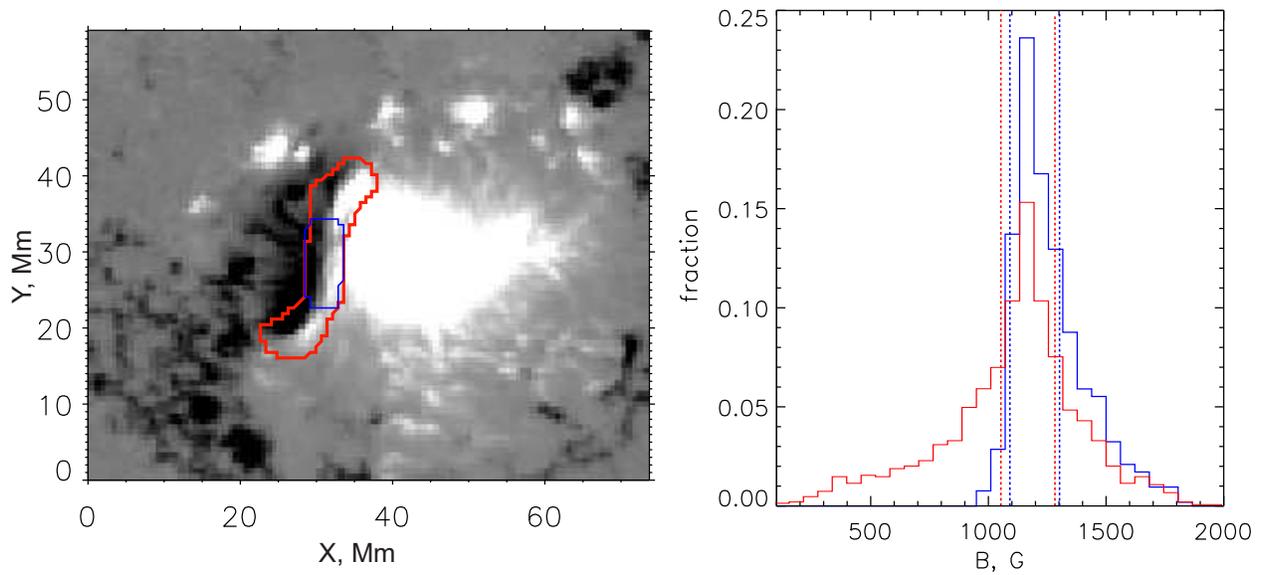}
\caption{Left panel shows map of $B_z$ component reprojected onto heliographic grid. Red and blue contours mark regions where distributions of magnetic field absolute values were calculated. The histograms of these two distributions are plotted in right panel and marked by corresponding colors. Vertical dotted lines mark FWHM levels for both distributions.}
\label{Bdistrib}
\end{figure}

The distribution of the magnetic field strength is presented in Fig.\,\ref{Bdistrib}. We selected two regions of interest (ROI) in the PIL to find the histogram of magnetic field distribution. The first ROI marked by red color in the left panel of Fig.\,\ref{Bdistrib} covers larger area than the other one (blue color). Distributions of the magnetic field are shown in the right panel of Fig.~\ref{Bdistrib}. One can see that the maximal probable values of these two distributions are approximately the same and equal to 1180~G. FWHM (marked by vertical dotted lines) is about 200 G for both distributions. That's why large number of the volume cells in the vicinity of the PIL have the values of the magnetic field in the range from 1100 up to 1300 G. The smaller ROI has the minimal value of the magnetic field around 1000~G, when the larger one reveals magnetic field values in the range of 300-1000~G.

Results of the magnetic field extrapolation in the flare region are imported to GX Simulator (see. subsection~\ref{GXres}). Magnetic field cube will be used for three-dimensional modelling of gyrosynchrotron radio emission generated by the nonthermal electrons transported in the found twisted magnetic structure elongated in the PIL.

%---------------------------------------------------------------------------------------------------------------------------------------------------------------------
    \subsection{GYROSYNCHROTRON RADIO EMISSION MODELLING FOR UNIFORM SOURCE}

In the previous sections it was shown that the accelerated electrons were injected into the compact magnetic structure elongated along the PIL (or they were directly accelerated there). NLFFF extrapolation of the magnetic field in the flare region reveals that magnetic field in the PIL does not significantly vary in space. As a first step, we have decided to make simple quantitative analysis of the flare microwave gyrosynchrotron spectrum assuming uniform source. The Stokes $I$ radio spectrum is calculated using the \textsf{Fast Gyrosynchrotron Codes} by \cite{Fleishman2010} where the authors used some analytical approaches and numerical methods to calculate the microwave emission with high speed and good accuracy for different energy and pitch-angle distributions of nonthermal particles. For simplicity the uniform pitch-angle distribution is considered. Angle $\Theta$ between line-of-sight and magnetic field is taken as 80~degrees. Parameters of the thermal plasma and spectrum of nonthermal electrons (at the HXR and microwave peak) were taken from the X-ray spectral fitting described in the Section~\ref{fitting}.

Four geometrical cases with different sizes of the region with nonthermal electrons are considered. Emitting region is assumed to be rectangular with the length $L$ comparable with the linear size of the observed magnetic structure parallel to the PIL, where nothermal electrons were transported. The width is equal to the line-of-sight depth $2R_{MW}$. We consider models with different $R_{MW}$ (0.15, 0.25, 0.5, and 1~Mm) and $L$ (13.3, 8, 10, and 10~Mm) presented in the Table~\ref{table1}. These values of $R_{MW}$ give different areas $2R_{MW}L$ of the emitting region in the plane of sky and cross section area $4R_{MW}^2$ (see table~\ref{table1}).

Such linear sizes of the emitting region in the four models were selected by the following reason. The scope is to obtain gyrosynchrotron spectrum peak frequency higher than 17~GHz and maximal observed radio flux. Figure~\ref{ModUniSource} demonstrates how peak intensity and frequency of gyrosynchrotron spectrum depend on density of nonthermal electrons and magnetic field strength in the source. The levels of constant peak frequency are marked by solid lines. The dotted line corresponds to the constant peak intensity with value of 350 sfu, that is maximal observed radio flux at frequency of 35~GHz during the studied flare (Fig.\,\ref{TP}b,\,c). Grey stripe marks range of the magnetic field values in the microwave source determined from the magnetic field distribution histogram obtained from the NLFFF extrapolation (Fig.\,\ref{Bdistrib}).

Considering intersection of the dotted and solid lines with the grey stripe one can deduce approximate density of nonthermal electrons $n_{MW}$ in radio source and peak frequency $f_p$ of the resulted gyrosynchrotron radio spectrum (see corresponding lines in table~\ref{table1}). In the case of $f_p = 35$~GHz we have $n_{MW}\sim 10^9$~cm$^{-3}$. Assuming cross section of the magnetic loop in the footpoints equal to $4R_{MW}^2$ one can deduce density of the precipitated nonthermal electrons producing HXR emission. From formula~\ref{nnth} taking $F(E>E_{low})=10^{35}$~electrons/s we have estimated the values of $n_{HXR}$ and compared them with $n_{MW}$ in the table~\ref{table1}. In the case of $R_{MW}=0.25$~Mm the value of $n_{HXR} = 4.6\times 10^9$~cm$^{-3}$ is the closest to the $n_{MW}=2\times 10^9$~cm$^{-3}$.

Using volume of the radio emitting region $V=4R_{MW}^2L$ we have estimated corresponding plasma density $n_{th} = \sqrt{EM/V}$ for $EM = 2.2\times 10^{46}$~cm$^{-3}$. For model~2 $n_{th} = 1.2\times 10^{11}$~cm$^{-3}$ and $n_{MW}/n_{th}\approx 0.017$. That is only 1.7~\% of electrons are accelerated from thermal plasma.

To make more precise analysis of the radio emission observed by the Nobeyama Radio Observatory one can make three-dimensional modelling of microwave emission from the nonthermal electrons spreading in the magnetic structure elongated along the PIL. In the next section we will describe such modelling using GX Simulator tool.

%\begin{table}
%\caption{  }
%\label{table 1}
%\begin{tabular}{ccccc}     % define the column alignment
                           % l: left, c: center, r: right
%  \hline                   % horizontal line
%$f_{max}$ & 35~GHz & 35~GHz & 25~GHz & 20~GHz \\
%  \hline
%$n_{MW}$, cm$^{-3}$ & $4\times 10^9$ & $3\times 10^9$ & $10^8$ & $10^7$ \\
%$R_{MW}$, Mm &  0.15 & 0.25 & 0.5 & 1 \\
%$S_{MW}=2R_{MW}L$, cm$^2$ & $7\times 10^{14}$ &  $2\times 10^{15}$ & $8\times 10^{15}$ & $3.4\times 10^{16}$ \\
%$R_{HXR}$, Mm & 0.46  & 0.53 & 2.9 & 9.2 \\
%$S_{HXR}=\pi R_{HXR}^2$, cm$^2$ & $6.6\times 10^{15}$ & $8.8\times 10^{15}$ & $2.7\times 10^{17}$ & $2.7\times 10^{18}$ \\
%$F_{MW}$, $10^{34}$ electrons/s & 1.5  & 3.1 & 0.4 & 0.2 \\
%$F_{MW}/F_{HXR}$ & 0.11 & 0.22 & 0.03 & 0.01 \\
%  \hline
%\end{tabular}
%\end{table}

\begin{table}
\caption{Summary of four models describing gyrosynchrotron microwave emission from the uniform source with different geometrical parameters and density of nonthermal electrons}
\label{table1}
\begin{tabular}{|c|cccc|}     % define the column alignment
                           % l: left, c: center, r: right
  \hline                   % horizontal line
Model \# &  1 & 2 & 3 & 4 \\
  \hline
$R_{MW}$, Mm &  0.15 & 0.25 & 0.5 & 1 \\
L, Mm        &  13.3 & 8    & 10  & 10 \\
$\pi R_{MW}^2$, cm$^2$ & $7.1\times 10^{14}$  & $2\times 10^{15}$ & $7.9\times 10^{15}$ & $3.1\times 10^{16}$ \\
$S_{MW}=2R_{MW}L$, cm$^2$ & $4\times 10^{16}$ &  $4\times 10^{16}$ & $10^{17}$ & $2\times 10^{17}$ \\
$V_{MW}=S_{MW}L$, cm$^3$ & $9.4\times 10^{23}$ &  $1.6\times 10^{24}$ & $7.9\times 10^{25}$ & $3.1\times 10^{25}$ \\
$f_{p}$ ($I_{max}=350$ sfu), GHz & 35 & 35 & 25 & 20 \\
$n_{MW}$, cm$^{-3}$ & $2\times 10^9$ & $2\times 10^9$ & $10^8$ & $1.3\times 10^7$ \\
$n_{HXR}$ $(S_{HXR}=S_{MW})$, cm$^{-3}$ & $1.3\times 10^{10}$ & $4.6\times 10^9$ & $1.2\times 10^9$ & $3\times 10^8$ \\
$n_{th}$ $(V_{SXR}=V_{MW})$, cm$^{-3}$ & $1.5\times 10^{11}$ & $1.2\times 10^{11}$ & $1.7\times 10^{10}$ & $2.7\times 10^{10}$ \\
  \hline
\end{tabular}
\end{table}

%-------------------------------------------------------------------------------------------------------------------------------------------
 %   \subsection{SELECTION OF THE MAGNETIC LOOPS FOR GYROSYNCHROTRON RADIO EMISSION MODELLING IN GX SIMULATOR}

%-------------------------------------------------------------------------------------------------------------------------------------------
    \subsection{GYROSYNCHROTRON RADIO EMISSION MODELLING IN GX SIMULATOR}
    \label{GXres}

3D modelling of the microwave radio emission is made using the \texttt{GX Simulator} package \cite{Nita2015}. This IDL-based program is an interactive tool allowing to select magnetic structures involved in the flare process using coronal extrapolations of the magnetic field. Thermal and nonthermal particle distributions along and across the selected magnetic structures are defined by the analytical expressions. The Stokes $I$ microwave maps are calculated using the \textsf{Fast Gyrosynchrotron Codes} by \cite{Fleishman2010}. In our modelling, the main task is to explain the emission at 17 and 34~GHz as we have imaging data for these frequencies and know exactly where this emission comes from. However, despite of absence of observations at lower frequencies we will also discuss the emission at frequencies below 17~GHz as well and will select the appropriate model to explain the whole microwave spectrum from 2 to 35~GHz observed by NoRP.

To explain dominating emission at 35~GHz one should suspect peak frequency in the range of 17-35~GHz. In the previous section it was shown that to explain the observed $f_p$ and the maximal radio flux of 350~sfu one need to assume population of nonthermal electrons with density of around $2\times 10^9$~cm$^{-3}$ uniformly distributed in the thin worm-like radio emission source with the width of 0.5~Mm and length of 10~Mm. The magnetic field in the source is 1180~G. Such simple analysis allowed us to find appropriate real magnetic structure from NLFFF extrapolation and distribution of nonthermal electrons within the found structure to reproduce the observed microwave spectrum.

\begin{figure}[ht]
\centering
\includegraphics[width=1.0\linewidth]{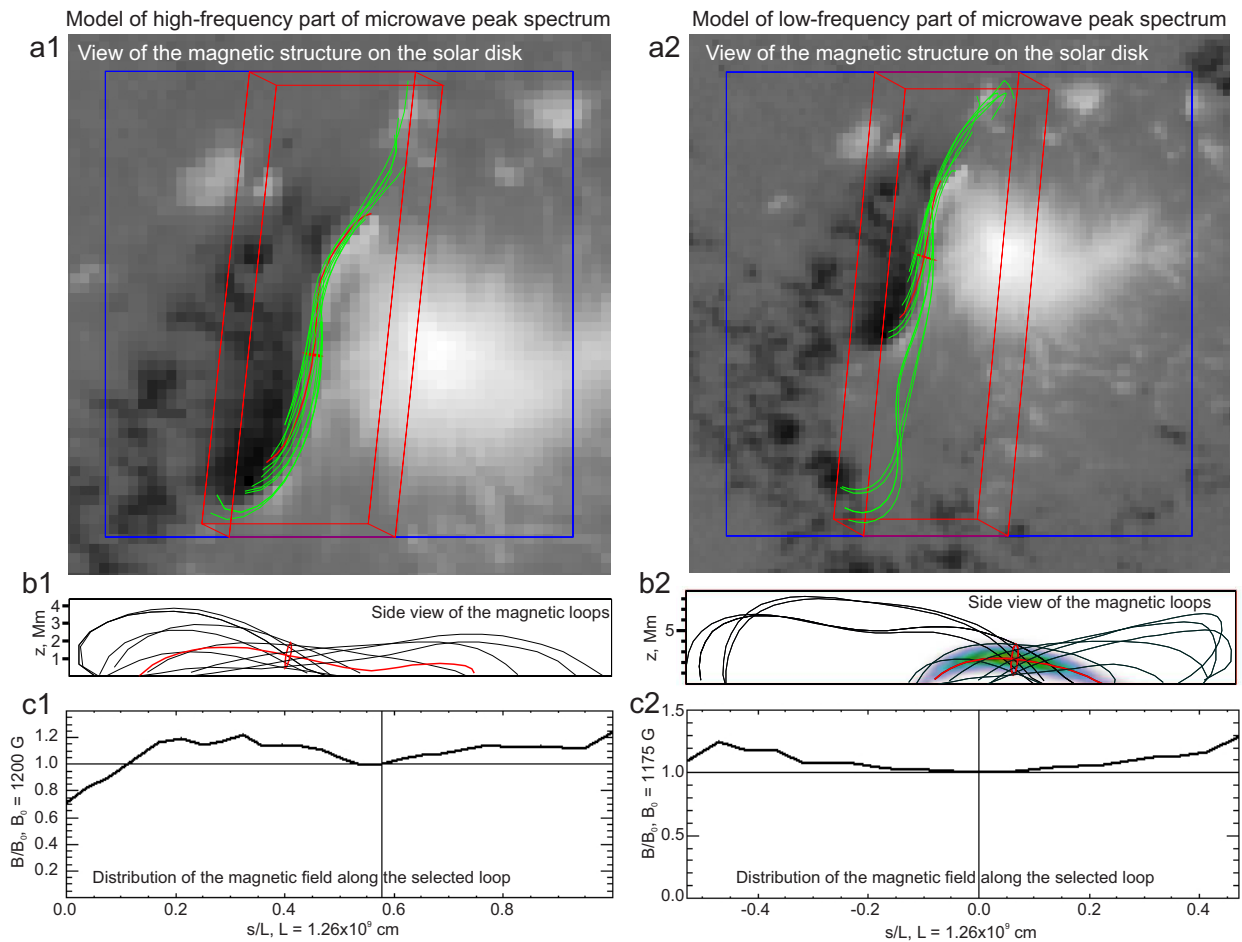}
\caption{Figure shows the selected magnetic structures in GX Simulator for microwave emission modelling. a) View of the selected magnetic structures on the solar disk with blue rectangle showing field of view in the plane of sky. Red lines correspond to the edges of the volume where the analyzed magnetic structure is located. b) Side view of the magnetic structure. c) Distribution of the magnetic field along the central line of the selected magnetic structure (also shown by red color). Left and right panels correspond to high-frequency (HF) and low-frequency (LF) models, respectively.}
\label{GXloops}
\end{figure}

\begin{figure}[ht]
\centering
\includegraphics[width=0.75\linewidth]{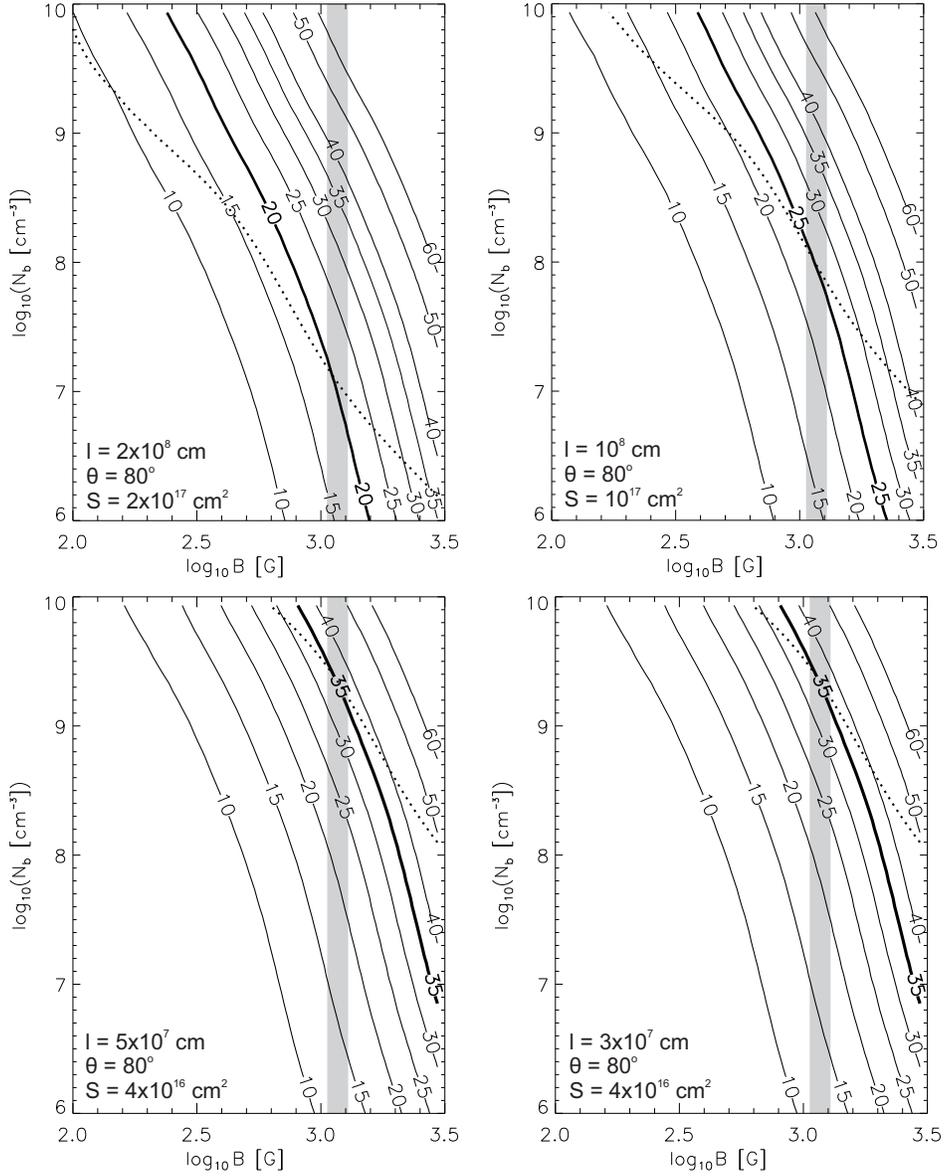}
\caption{Figure presents plots where solid lines mark levels of constant peak frequency of the gyrosynchrotron radio spectrum. Dashed line corresponds to the level of constant radio flux of 350~sfu that was maximal during the studied flare. Values of magnetic field and density of nonthermal electrons are written in the X and Y axes. Grey stripe shows FWHM of magnetic field distribution in the PIL obtained from the histograms shown in right panel of Fig.\,\ref{Bdistrib}. Four panels correspond to different sizes of the region emitting gyrosynchrotron microwave emission (see table~\ref{table1}).}
\label{ModUniSource}
\end{figure}

As we previously mentioned, we have imported results of NLFFF extrapolation into GX Simulator. We have selected the magnetic structure using selection tool in GX SIMULATOR by the following way. Firstly we defined central line of the loop to achieve the closest correspondence between the central line location and HXR footpoints and microwave sources at available NoRH working frequencies of 17 and 34 GHz. Then, we defined circular cross section of the magnetic structure at the top of the central line. The cross-section radius $R$ of the loop is 0.72~Mm and length of the central line $L = 12.6$~Mm. To visualize magnetic structure 12 lines were drawn from the points distributed along the circle around top point of the central line. The magnetic structure is presented in Fig.\,\ref{GXloops}a1,\,b1 in two projections: on disk view and side view. This structure is elongated along the PIL that is in accordance with the observations made in different available ranges of electromagnetic spectrum. The top point of the central line has height of $\approx 1.5$~Mm. Thus, the flare magnetic structure is low-lying and located in the chromosphere or just slightly above it.

Nonthermal electrons are non-uniformly distributed inside the magnetic structure. We consider a Gaussian shape of the distributions of the nonthermal electrons along and across the magnetic structure. The distributions are determined by the expressions $n(r,l) = n_0\exp{[-(3r/R)^2]}\exp{[-(3l/L)^2]}$, where $R$ is the loop radius, $r$ is the coordinate across the loop, $L$ is the loop length, and $l$ is the coordinate along the loop from its top point. The peak number density is chosen to be $n_0=5\times 10^8$ $\textrm{cm}^{-3}$ at the loop top.  The efficient length of the radio source $L_{MW} = L/3 = 4.2$~Mm when $R_{MW} = R/3 = 0.24$~Mm.

Distribution and density of nonthermal electrons in the loop were chosen to reproduce the observed radio source and obtain a sufficient radio flux at 17 and 34~GHz to fit the radio spectrum measured by NoRP. The energy spectrum of nonthermal electrons obtained from the analysis of X-ray spectrum is described by power-law function with spectral index $\delta=3.5$, the low-energy cutoff $E_{\mathrm{low}}=20$ keV, and the high-energy cutoff $E_{\mathrm{high}}=10$ MeV. For simplicity, we consider an isotropic pitch-angle distribution of the nonthermal electrons. The background thermal plasma is uniformly distributed in the loop and has the number density $n_{\mathrm{th}} = 10^{11}$~$\textrm{cm}^{-3}$ and temperature of $T=30$~MK. Previous analysis assuming uniform source reveal  that nonthermal electrons density of $2\times 10^9$~$\textrm{cm}^{-3}$ is enough to explain radio spectrum. Using GX SIMULTOR we found lower density (by a factor of 4) as we considered magnetic structure obtained from the NLFFF extrapolations with the non-uniform magnetic field and changing orientation (inclination to line-of-sight) in the space.

\begin{figure}[ht]
\centering
\includegraphics[width=0.5\linewidth]{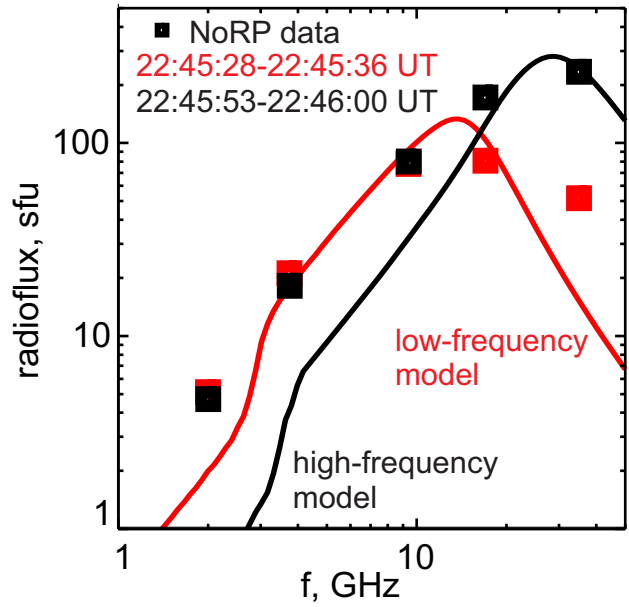}
\caption{Comparison of the two NoRP spectra with the model spectra from GX Simulator made for LF and HF cases (see Fig.\,\ref{GXloops}). NoRP spectra are the same as those ones shown in the first two columns of Fig.\,\ref{spec}. LF and HF model spectra are marked by red and black colors, respectively.}
\label{model_spec}
\end{figure}

We have just described the model of high-frequency part of the microwave spectrum. However, one can see (Fig.\,\ref{model_spec}) that this model fails to explain the whole wide radio spectrum. Below, we will describe a way to explain the emission at frequencies of 2, 3.75, and 9.4~GHz. Thus, we will discuss low-frequency model (hereinafter LF model). For construction of LF model we selected a bit higher and wider magnetic structure in the vicinity of the PIL. It is shown in Fig.\,\ref{GXloops}a2,\,b2. Distribution of the magnetic field along the loop is presented in panel c2. The radius of the magnetic structure is $1.44$~Mm, length of the central line is 17.6~Mm. This magnetic structure is filled by nonthermal electrons distributed along and across the central line according to the following formula $n(r,l) = n_0\exp{[-(1.6r/R)^2]}\exp{[-(2.2l/L)^2]}$. The efficient length of the radio source $L_{MW} = L/2.2 = 8$~Mm when $R_{MW} = R/1.6 = 0.9$~Mm. The density at the loop top was taken as $n_0 = 10^7$ $\textrm{cm}^{-3}$. Such density distribution was selected to achieve sufficient radio flux to explain the low frequency (below 17 GHz) part of the microwave spectrum. Power-law index, low and high energy cutoffs were selected the same as in the case of HF model suggesting the same accelerator of electrons. The model spectrum is shown by red solid line in Fig.\,\ref{model_spec}. The peak frequency of LF spectrum is about 13~GHz.

Geometrically the constructed model can be described by the following way. There is a thick (large width) magnetic loop with low-density population of nonthermal electrons with high-density beam of nonthermal electrons spreading in a thin channel inside thicker one. The low- and high-density populations of nonthermal electrons produce low- and high-frequency radio emissions, respectively. As a result, one can see that the whole spectrum is nicely fitted by the combination of LF and HF models. However, the constructed LF model is rather speculative because there is no any images at the frequencies below 17~GHz. Thus, we cannot localize magnetic structure from where radio emission at lower frequencies are emitted. Here, we make assumption that radio emission at low frequencies is also emitted from the PIL region, as all strongest emission sources were close to the PIL. However, we think that it is possible to create another LF model by playing with geometry of the magnetic structure and distribution of the nonthermal electrons inside it. Anyway, we have demonstrated that electrons in the closed twisted magnetic structure in the PIL can produce broad-band gyrosynchrotron microwave emission. Emission at 17 and 34~GHz is definitely generated by high-density beam of the nonthermal electrons transported in very thin low-lying magnetic channel in the PIL.

%%%%%%%%%%%%%%%%%%%%%%%%%%%%%%%%%%%%%%%%%%%%%%%%%%%%%%%%%%%%%%%%%%%%%%%%%%%%%%%%%%%%%%%%%%%%%%%%%%%%%%%%%%%%%%%%%%%%%%%%%%%%%%%%%%%%%%%%%%%%%%%%%%%%%%%%%%%%%%%%%%%%%%
\section{DISCUSSION AND CONCLUSIONS}

In this work we presented detailed analysis of the spatially-resolved observations of the M1.2 solar flare occurred on March 15, 2015. This event was selected for the analysis due to its strong emission sources in the PIL pointing to interaction of stressed magnetic loops with the high shear angle (up to 80 degree) experienced three-dimensional magnetic reconnection. Another reason to select this event was the good association of the HXR sources with the regions of strong photospheric vertical electric currents. It means that accelerated electrons were indeed accelerated/injected in the twisted magnetic structure. Our analysis allowed to determine physical parameters of accelerated electrons, heated plasma and magnetic field topology in the PIL where the observed emission sources were localized. The main results can be summarized in a following way:

\begin{enumerate}

\item Accelerated electrons and heated plasma were localized in the closed low-lying twisted magnetic structure in the PIL with the average height of up to 3~Mm. The average magnetic field value was about 1200~G.
\item The most energetic electrons with hardest spectrum were appeared at the onset of plasma heating up to super-hot temperature of 40~MK. Plasma beta in the super-hot region was less than 0.01.
\item The density of the accelerated electrons in the PIL was about $10^9$~cm$^{-3}$. Estimations show that less than 10~\% of electrons are accelerated from the thermal super-hot plasma assuming the same location of thermal and nonthermal populations.
\item The largest part of the total flare energy release during the impulsive phase was concentrated in the accelerated electrons. Total kinetic energy of the accelerated electrons was about $10^{30}$~ergs when the thermal energy of the super-hot plasma did not exceed $3\times 10^{28}$~ergs. Nonthermal electron energy flux is estimated up to $3\times 10^{12}$~ergs~s$^{-1}$~cm$^{-2}$, that is much higher than in the currently available flare radiative hydrodynamic models.
\item Joint analysis of the HXR, microwave, EUV and optical data revealed that the accelerated electrons were transported in a thin magnetic channel within a twisted magnetic structure. The width of this channel was about 0.5~Mm when the total area of optical flare ribbons was up to $2\times 10^{17}$~cm$^2$. Thus, only a part of the twisted magnetic structure was involved in the process of efficient (up to high energies) acceleration.

\end{enumerate}

It was found that the flare energy release was developed in the low-lying twisted magnetic structure elongated along the PIL. The most possible magnetic configuration for the initial flare energy release is tether-cutting magnetic reconnection, where magnetic loops with the high shear interacts in the PIL and experience three-dimensional magnetic reconnection with the strong guiding magnetic field of $\sim 1$~kG. In the studied flare, these interacting magnetic structures are located in the twisted MFR found from the NLFFF extrapolation of magnetic field. During the flare energy release nonthermal electrons were injected into the very thin magnetic channel comparing with the observed area of the flare ribbons. Thus, the acceleration process is connected with filamentation in the reconnecting sheared magnetic loops within the MFR. It should be also noticed that the previous works reported mainly the TCMR during solar flares in large scale coronal magnetic structures \citep[e.g.][]{Liu2013}. Possibly in such low-lying magnetic structures influence of partially ionized plasma and large gradients (chromosphere-corona interface) on magnetic reconnection may play a role. Future models should be able to reproduce complex physics of three-dimensional magnetic reconnection in the PIL involving different atmospheric layers and producing accelerated electrons and super-hot plasma.

Importance of fine spatial structuring of the flare energy relase site was previously discussed in the recent works of \cite{Krucker2011,Zimovets2013,Sharykin2014a,Yurchyshyn2017}, where authors used observation with high spatial resolution in different ranges of electromagnetic spectrum. We think, the important result of this study is the filamentation of flare energy release with subsequent formation of high energy density beams of nonthermal electrons. It should be again noticed that numerical modeling of gas dynamics response in flare regions previously did not consider such energy densities of $3\times 10^{12}$~erg~s$^{-1}$~cm$^{-2}$. Numerical radiative gas dynamics models should be able to reconstruct behavior of flare chromospheric plasma under such energy fluxes. It is important for explanation of lower solar atmosphere response to injection of nonthermal electrons. For example, in the work of \cite{Sharykin2017a} it was discussed that high density beams of nonthermal electrons can be a reason for white-light emission and sunquake generation. Moreover, dense population of nonthermal electrons will lead to very strong induced return currents and more efficient interaction with plasma waves and generated turbulence. These effects are should be also taken into account in the kinetics of accelerated electrons in the solar flares. However, despite on large energy density of nonthermal electrons, our analysis revealed that the super-hot plasma has enough particles to be accelerated. Ideally, self consistent model of three-dimensional magnetic reconnection in the PIL with particle acceleration and their kinetics should be constructed. Such model has to reproduce formation of high energy density nonthermal electrons population in the contest of filamentation of flare energy release site in the MFR located in the PIL.

One of the interesting finding in this work is the formation of the super-hot plasma with temperature reaching value of 40~MK during the flare. Great interest to this phenomenon is connected with the fact that such large temperature can be resulted from direct plasma heating in the primary energy release site in the solar corona \citep{Caspi2010}. In our case the super-hot plasma was formed in the PIL during probable tether-cutting magnetic reconnection in the low-lying MFR. Thus, numerical models have to reconstruct such extreme heating and low plasma beta. According to the statistical work of \cite{Caspi2014} it was shown that the plasma beta in the super-hot region cannot be much less than unity. However, in our work, the estimated average magnetic field is about 1200~G, that is very high, and the resulted plasma beta in the super-hot region is less than 0.01. Thus, the super-hot plasma is fully magnetized in contrast with the results of \cite{Caspi2010, Caspi2014, Sharykin2015}. Here we need to note that the high value of the magnetic field found in the flare region can be explained by the fact the flare energy release happened in the very low-lying (just up to 3~Mm) magnetic structure near the sunspot. This flare was not accompanied by an eruption and a CME, i.e. the flare was confined and did not develop into a normal eruptive event.

Another peculiarity is that the total energy of nonthermal electrons is much larger than the internal energy of the super-hot plasma. In the work of \cite{Sharykin2015} the kinetic power of nonthermal electrons was comparable with the time derivative of the super-hot plasma internal energy. However, in the work of \cite{Sharykin2014}, it was shown that the kinetic power of nonthermal electrons can dominate over thermal energy. The conclusion was that acceleration can result in effective cooling due to efficient escape of fast electrons from the Maxwellian tail of the super-hot plasma. In this work we found two orders magnitude difference between thermal and nonthermal energies. It seems that magnetic reconnection in the low-lying magnetic loops is very efficient accelerator but not efficient heater from the energy point of view. To sum up, the largest part of the flare energy release is concentrated in the accelerated electrons, but the change of magnetic free energy in the PIL region is enough to explain the total flare energetics, that is consistent with the previous studies \citep[e.g.][]{Emslie2012}.

Finally, it is worth noting that this work (Paper~I) was mostly devoted to investigation of plasma heating and nonthermal electrons in the PIL region. We ignored here the detailed investigation of magnetic field dynamics in the PIL. The subsequent work called ``Paper~II'' will present the detailed study of magnetic fields, electric currents and their relation to emission sources in the PIL using the high-cadence HMI vector magnetograms \citep{Sun2017} with temporal resolution of 135~seconds.

\acknowledgements

We are grateful to the teams of RHESSI, HMI/SDO, AIA/SDO, Nobeyama Solar Radio Observatory, SOT/Hinode for the available data used in this study. This work is supported by the Russian Science Foundation under grant 17-72-20134.

% The typical value of the magnetic field in the super-hot region is about 100-300~G. In our flare the maximal temperature of the super-hot region was about 40~MK.

% \item Accelerated electrons were transported in the magnetic field structure with average magnetic field value of 1200~G.

%%%%%%%%%%%%%%%%%%%%%%%%%%%%%%%%%%%%%%%%%%%%%%%%%%%%%%%%%%%%%%%%%%%%%%%%%%%%%%%%%%%%%%%%%%%%%%%%%%%%%%%%%%%%%%%%%%%%%%%%%%%%%%%%%%%%%%%%%%%%%%%%%%%%%%%%%%%%%%%%%%%%%%

\bibliographystyle{apj}
%\bibliography{bibl}

\clearpage
%%%%%%%%%%%%%%%%%%%%%%%%%%%%%%%%%%%%%%%%%%%%%%%---------Figures-----------%%%%%%%%%%%%%%%%%%%%%%%%%%%%%%%%%%%%%%%%%%%%%

%_____________________________________OBSERVATIONS__________________________________________

%----------------- time profiles of RHESSI and Nobeyama data ---------------------------------

%----------------- jz compare with Hinode and RHESSI ---------------------------------

%----------------- NoRH images and the PIL ---------------------------------

%----------------- AIA images, RHESSI ims and the PIL ---------------------------------

%----------------- Hinode images, RHESSI and the PIL ---------------------------------

%____________________________RHESSI SPECTRAL ANALYSIS_________________________________

%----------------- RHESSI and NoRP spectra ---------------------------------

%----------------- RHESSI fit results ---------------------------------

%----------------- Energetics ---------------------------------

%____________________________MW analysis results_________________________________

%----------------- Magnetic field distribution in the PIL ---------------------------------

%----------------- GX_Simulator Magnetic structure ---------------------------------

%----------------- Uniform source MW modelling ---------------------------------

%----------------- GX_model spectra ---------------------------------
%---------------------------------------------------------------------

\clearpage
\end{document}